\documentclass[
 preprint,prx,
 amsmath,amssymb]{revtex4-2}

\usepackage{CJK}
\usepackage{graphicx}
\usepackage{dcolumn}
\usepackage{bm}
\usepackage{float}
\usepackage{color}

\usepackage{enumerate}

\begin{document}
\begin{CJK*}{}{} 
\author{Matteo De Tullio,$^{1\ast}$ 
Giovanni Novi Inverardi,$^{2}$ 
Michella Karam,$^{1}$ 
Jonathan Houard,$^{1}$ 
Marc Ropitaux,$^{3}$ 
Ivan Blum,$^{1}$ 
Francesco Carnovale,$^{2}$ 
Gianluca Lattanzi,$^{4}$ 
Simone Taioli,$^{5}$ 
Gustav Eriksson,$^{6}$ 
Mats Hulander,$^{6}$ 
Martin Andersson,$^{6}$ 
Angela Vella$^{1\ast}$
Tommaso Morresi$^{5\ast}$\\~\\}
\affiliation{$^{1}$Universit{\'e} Rouen Normandie, INSA Rouen Normandie, CNRS, GPM UMR 6634, F-76000 Rouen, France\\
$^{2}$Department of Physics, University of Trento and European Centre for Theoretical Studies in Nuclear Physics and Related Areas (ECT*-FBK) and Trento Institute for Fundamental Physics and Applications (TIFPA-INFN), Trento, Italy\\
$^{3}$Universit{\'e} Rouen Normandie, GLYCOMEV UR4358, SFR Normandie V{\'e}g{\'e}tal FED 4277, Innovation Chimie Carnot, IRIB, F-76000 Rouen, France\\
$^{4}$Department of Physics, University of Trento and Trento Institute for Fundamental Physics and Applications (TIFPA-INFN), Trento, Italy\\
$^{5}$European Centre for Theoretical Studies in Nuclear Physics and Related Areas (ECT*-FBK) and Trento Institute for Fundamental Physics and Applications (TIFPA-INFN), Trento, Italy\\
$^{6}$Department of Chemistry and Chemical Engineering, Chalmers University of Technology Gothenburg 41296, Sweden
}%
\title{Evaporation of cations from non-conductive nano-samples using single-cycle THz pulses: an experimental and theoretical study}

\begin{abstract}
This study investigates the emission of cations from silica samples by single-cycle THz pulses, focusing on the influence of pulse polarity. 
Negative THz pulses were found to efficiently trigger the evaporation of cations from nanoneedles in amorphous silica samples compared to positive pulses. Conversely, this dependence on pulse polarity could not be found in samples with metallic behaviour such as LaB$_6$ and when multi-cycle pulses in different frequency ranges such as ultraviolet (UV) are used.
First-principles simulations focus on silica under THz laser irradiation and show critical fields for ion evaporation of hydroxyl groups from Si(OH)$_4$, which serves as a model precursor molecule for the amorphous solid matrix. To explain our experimental results, we propose a simplified theoretical model that determines the role of the polarity of the THz pulse by taking into account the differences in electron mobility between silica and semi-metallic samples. The study explores the nonlinear microscopic mechanisms of atomic evaporation under external static and THz laser fields and clarifies the dynamics of THz-enhanced APT and related applications.
\end{abstract}


\maketitle
\end{CJK*}
\onecolumngrid
The interaction of electromagnetic fields with nanostructures is attracting increasing interest for various applications including imaging, ultrafast electron microscopy, gas sensing and controlled catalysis \cite{sadeghian2011ultralow,liu2007field,ovchinnikov2019second, klarskov2017nanoscale,dombi2020strong,larue2015thz}. \\
\indent In the last ten years, this has led to a surge in research activities focusing on the manipulation of materials by external fields. Two main approaches have emerged: leveraging ultra-strong electrostatic fields and employing fast and powerful optical pumping \cite{silaeva2014dielectric, schultze2013controlling, herink2012field}.  \\
\indent
In particular, in terahertz science, novel techniques for generating intense and broadband pulses have been developed, ranging from optical rectification in nonlinear crystals to two-colour plasma generation in air/gas/liquid using femtosecond lasers \cite{dai2009coherent, sato2013terahertz, dechard2018terahertz, koulouklidis2020observation}.
THz pulses bridge the gap between these two directions of controlling matter through static or electromagnetic fields.
High-intensity THz pulses are usually generated with large free-electron laser systems such as the Linac Coherent Light Source \cite{larue2015thz} or by the interaction of low-amplitude THz pulses with metallic nanostructures. In particular, metallic nanoneedles were used to amplify the THz field, enabling the emission of electrons and ions from their surface \cite{wimmer2014terahertz, li2016high, vella2021high, karam2023thz}.
Picosecond duration THz pulses are suitable for many applications such as the activation and control of catalytic reactions \cite{larue2015thz} and, recently, their ability to trigger electron and ion emission from nanometric samples has been demonstrated \cite{wimmer2014terahertz, li2016high,vella2021high,karam2023thz}.\\
\indent These sources show the potential for ultrafast manipulation of matter and dynamics at interfaces, including materials/vacuum and solid/liquid interfaces \cite{jelic2017ultrafast,herink2012field, li2014ionization, huzayyin2014interaction}. 
Single-cycle THz pulses with high amplitudes are used in imaging techniques such as scanning near-field microscopy (SNOM) and atom probe tomography (APT) and enable the analysis of materials with different properties. \\
\indent 
The shape of the THz pulse, in particular its polarity (positive, negative or bipolar), can significantly influence the interaction between THz light and material, especially with regard to the emission of positive or negative charges. However, the interaction of THz transients with non-conducting nanostructures and their ability to emit electrons or ions, as well as the influence of the pulse shape on the ion emission process, have not yet been investigated to our knowledge.\\
\indent In this work, we study the interaction of THz single-cycle pulses with nanoneedles made of amorphous silica, a material with a large band gap of about 9 eV and low electrical conductivity. We conduct experimental analyses on ion emission in an APT chamber and complement these with numerical analyses using time-dependent density functional theory (TDDFT) method.\\
\indent 
In experiments, THz single-cycle pulses with positive and negative polarity are focussed on the sample. Previous research has demonstrated THz-induced ion evaporation from highly conductive materials such as aluminium and lanthanum hexaboride (LaB$_6$) using positive THz pulses \cite{vella2021high,karam2023thz,karam2024thz}. In this study, we present the novel finding that negative THz pulses can also trigger the evaporation of cations from nanoneedles by comparing the results from silica and LaB$_6$ samples. Negative pulses are found to be more effective than positive pulses, especially for amorphous silica samples.\\
\indent
On the experimental side, this study presents for the first time the single-cycle evaporation of cations from non-conducting nanosamples using THz pulses. It emphasises the importance of pulse polarity for the emission process. These findings have significant implications for the further development of THz-assisted APT, THz-SNOM and other configurations where nanostructures are used to amplify the THz field and induce molecular ionisation, such as THz photochemistry and THz-induced surface reactions.\\
\indent From a theoretical perspective, the investigation of THz-based ion field emission requires an explicit consideration of the electronic degrees of freedom. We also emphasise the importance of a time-dependent approach to capture the nonlinear effects associated with the interaction of the sample with the THz source and the static electric field, leading to dynamic changes in the interaction forces within the sample.\\
\indent
TDDFT \cite{ullrich} can be combined with molecular dynamics (MD) in the framework of Ehrenfest dynamics \cite{marx_2009} to study this phenomenon. However, the high computational cost of TDDFT to achieve the accuracy required to describe evaporation processes limits the size of the simulated systems. Furthermore, the evaporation of cations from a solid sample is a very complex process that cannot be realistically treated using ab initio simulations. This is due to competing processes that contribute to the breaking of bonds on the sample surface, such as ionisation and thermal effects. The latter are very difficult to model and disentangle in first-principles dynamic simulations. Therefore, here we mainly focus on the role that ionisation plays in atomic evaporation.
Currently, there are very few computational studies using TDDFT to explore THz-induced ion field emission, with some focussing on UV femtosecond laser pulses \cite{Silaeva_2015} or static methods \cite{feibelman_2001, sanchez_2004, tamura_2012}. Here we present a TDDFT-based model to investigate the behaviour of a silicon dioxide matrix under an electrostatic field and THz laser irradiation using Si(OH)$_4$ as a model molecule. Si(OH)$_4$ is the precursor molecule of the amorphous solid matrix synthesised in this work using the sol-gel method for the encapsulation of biomolecules \cite{sundell_2019}.
This approach enables the identification of critical THz laser fields and mechanisms for the ion evaporation process of the hydroxyl groups of Si(OH)$_4$.\\
\indent 
However, analysing a single molecule is limited by the lack of a band structure and the size constraints of the system.
In a realistic sample, electrons are either confined or free to move, depending on whether it is an insulating or metallic material. Therefore, we introduce a simplified toy model that accounts for the different electron mobility of silica compared to semi-metallic samples such as LaB$_6$ to describe and analyse the different behaviours observed experimentally. This model allows us to study the different regimes of static and laser fields and to understand the role of ionisation in the evaporation of cations from solid samples, as in APT experiments. The dependence of ion emission on pulse polarity, which was observed experimentally in the THz domain, can also be predicted theoretically in the visible IR range, so that the conclusions of this study can be extended beyond the THz range to all spectral ranges in which high-intensity single-cycle pulses can be generated \cite{hwang2019generation}.

\section{Experimental and theoretical methods}

\subsection{Experimental methods}

For the LaB$_6$ samples, LaB$_6$ single crystal rods with a diameter of 0.7 mm and a length of 10 mm, orientated along the $<100>$ direction, were obtained directly from APTech, USA. The tip of the LaB$_6$ rod was electrochemically etched in a 10\% HNO$_3$ solution to obtain a sharp apex. Further refinement of the tip to achieve a tip diameter of less than 20 nm was performed with a dual-beam focussed ion beam scanning electron microscopy (FIB-SEM) system (Helios 5 Plasma FIB) using Xe ions. The majority of the milling process was performed with 30 keV ions, while a final milling step at 12 keV was used to minimise the presence of defects in the resulting sample \cite{blum2016atom}.\\
\indent 
For the silica samples, the specimens were prepared according to a protocol developed for the encapsulation of proteins in silica for APT analysis \cite{sundell2019atom} whose early stages were studied by molecular dynamics \cite{noviinverardi_2023}. In all experimental steps, the water used was filtered through a Milli-Q (Millipore) system. All chemicals were purchased from Sigma-Aldrich and used without further purification. A sodium silicate solution (25.5 - 28.5 \% SiO$_2$, 7.5 - 8.5 \% Na$_2$O), also known as water glass, was used to form silica through a sol-gel process. Gelling was initiated by adjusting the pH of the sodium silicate solution from alkaline to physiological (pH 7.4). For this purpose, an acidic ion exchange column was prepared by adding 1 g of ion exchange resin (Dowex 50WX8 hydrogen form 50-100 mesh, Sigma-Aldrich) to 0.5 g of glass wool in a 10 mL plastic syringe. The ion exchange resin was activated by washing with 10 mL NaOH (4 M), followed by water, and finally 10 mL HCl (1 M). The activity was adjusted by rinsing with small amounts of water until the eluate had a neutral pH. The original sodium silicate solution was then diluted with water in a volume ratio of 1:3 before 1 mL of the solution was passed through the column, collected in a vial and allowed to harden in an open vial at 37$^\circ$ C for 48 h to obtain a solid silicate material. The silica samples were prepared using a lift-out method with a FIB-SEM dual-beam system (Helios 5 Plasma FIB), in which Xe ions were used to cut the matrix into a thin needle shape with a radius of less than 50 nm \cite{blum2016atom,thompson2007imaging}.
\begin{figure}[!htb]
   \centering
\includegraphics[width=1\linewidth]{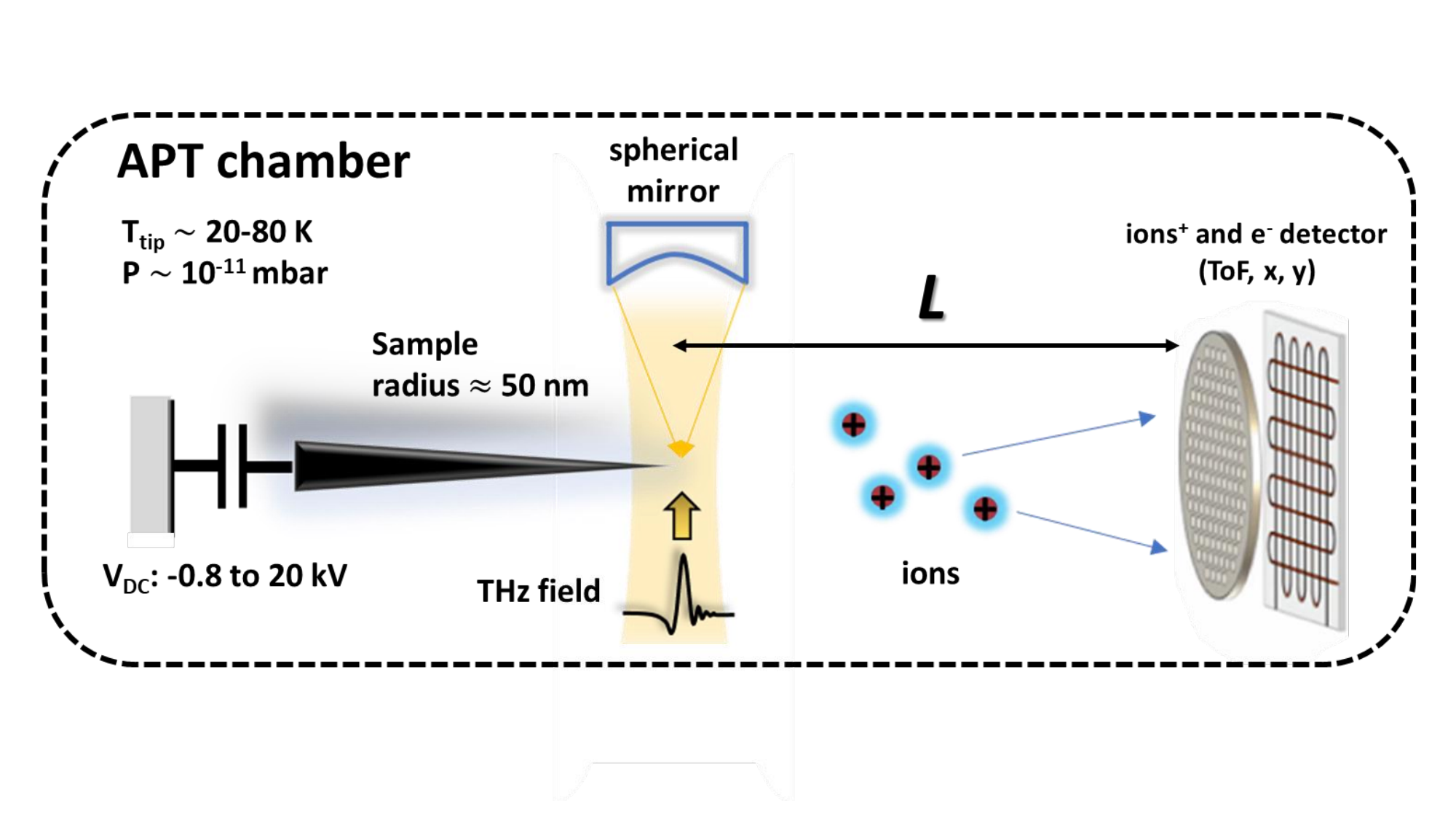}\caption{\label{fig:figure1} Experimental layout: Ultrafast terahertz pulses (blue line) are focussed on a nanotip in a high-vacuum chamber. The high voltage applied to the metal tip leads to an intense electric field at the apex of the sample. The evaporated ions are projected onto a time- and position-sensitive detector located at L=10 cm away from the nanotip.}
\end{figure} 
The spatially resolved mass spectrum is measured with a THz-assisted APT system, the layout of which is shown in Fig. \ref{fig:figure1} \cite{blavette1993atom}. The nanometric needle-shaped specimen is polarized with a DC voltage ranging from -0.8 kV to 20 kV, resulting in an apex electric field exceeding 10 V/nm. The atoms in the sample layers are evaporated by the strong electrostatic field and irradiation with single-cycle THz pulses, which initiate the evaporation process \cite{vella2021high, karam2023thz}.\\
\indent 
Single-cycle THz pulses are generated in air by a two-colour laser field-induced plasma \cite{bartel2005generation, kim2007terahertz}. These THz pulses have a repetition rate of 1 kHz and are synchronised with the two-colour laser system, which comprises the fundamental harmonic (FH) and the second harmonic (SH) of a laser with a wavelength of 800 nm. The amplitude of the THz pulses is directly related to the input power of the FH, which can be adjusted at the plasma entry. The THz fields are mainly generated by asymmetric electron currents induced by photoionisation in air gas. The generated THz field is characterised by electro-optical (EO) scanning in a non-linear crystal before it is focused on the APT sample. The THz pulse with an electric field of 20 MV/m is amplified a thousandfold at the apex of the sample, as shown in previous studies on metallic and LaB$_6$ samples \cite{houard2020, karam2023thz}. Consequently, the maximum THz field amplitude is between 5-20 V/nm.\\
\indent 
After evaporation, the ions are directed onto a detector with the aid of an electrostatic field. The detector provides information about the impact position and the time of flight of the ions \cite{costa2012advance} enabling the reconstruction of the 3D structure of the sample. The chemical nature of each detected ion is determined by calculating the mass-to-charge ratio, as shown in the mass spectrum in Fig. \ref{fig:figure1a}(a) and each detected ion is assigned an initial position through back projection \cite{blavette1993atom} as shown in the mass spectrum in Fig. \ref{fig:figure1a}(b). 

\subsection{Theoretical and computational methods}

The coupled electronic-nuclear of Ehrenfest dynamics is formulated as follows (the atomic units are adopted) \cite{ullrich}:
\begin{equation}
    \begin{split}
i \frac{\partial \phi_j (\mathbf{r},t)}{\partial t} &= \Big[ -\frac{\nabla^2}{2} + v_{\mathrm{ext}}(\mathbf{r},t) +  v_{\mathrm{H,xc}}(\mathbf{r},t) + \hat{W}_{en} \left( \{ \mathbf{r} \} , \{ \mathbf{R}(t) \} \right) \Big] \phi_j(\mathbf{r},t)  \\
M_J \frac{\partial^2 \mathbf{R}_J}{\partial t^2} &= - \nabla_{\mathbf{R}_J} \Big[ v_{\mathrm{ext}} (\mathbf{R}_J,t) + \hat{W}_{nn} \left( \{ \mathbf{R}(t) \} \right) + \int d^3r \ n(\mathbf{r},t) \hat{W}_{en} \left( \{ \mathbf{r} \} , \{ \mathbf{R}(t) \} \right) \Big]
    \end{split}
\end{equation}
where $\phi_j$ is the electronic wave function of the occupied orbitals, $v_{\mathrm{ext}}$ is external scalar potential acting on both electrons at position $\{ \mathbf{r} \}$ and nuclei at position $\{ \mathbf{R} \}$, $v_{\mathrm{{H,xc}}}$ is the Hartree plus exchange-correlation potential, $\hat{W}_{nn}$ is the Coulomb interaction between the nuclei, $M_J$ is the mass of the $J$-th nucleus, $\hat{W}_{en}$ is the Coulomb interaction between the electrons and the nuclei, and $n(\mathbf{r},t) \equiv \sum_{j}^{occ} | \phi_j(\mathbf{r},t) |^2$ is the electron density.\\
\indent 
The simulation box for the Si(OH)$_4$ molecule is 20 \AA$^3$, chosen to conserve the system's energy without time-dependent fields. The lattice spacing is 0.16 \AA\ in each dimension, ensuring total energy convergence below 5 meV.
To prevent laser-emitted electrons from reflecting at the box edges, we apply absorbing boundary conditions that suppress electronic density. A mask function, equal to one in the inner region and smoothly decreasing to zero at the edges, is used for this purpose. 
The time step integrator is $4\cdot 10^{-4}$ fs, and ion trajectories and electron wave functions are propagated for 100 fs over 2.5$\cdot$10$^5$ steps.\\
\indent 
In the model system, we used only one dimension (1D), a box size of 80 \AA\ with a mesh spacing of 0.05 \AA, a time step of $5\cdot 10^{-4}$ fs for the time integration of the equations of motion, and a total of 3.2$\cdot$10$^5$ steps to track the dynamics of the system even after the THz field is removed.\\
\indent 
The initial geometries of Si(OH)$_4$ are shown in Fig. \ref{fig:SiOH4_geom}. They are optimized using the FIRE algorithm \cite{Bitzek_2006} with a force threshold of 0.03 eV/\AA. To prevent global translations and rotations of the molecule under the static external field, we clamped the central Si atom, the $y$ and $z$ coordinates of the oxygen atom O$1$, and the $y$ coordinate of the hydrogen atom H$1$ (see Fig. \ref{fig:SiOH4_geom}) during the optimization. The static electric field values considered in this work are $\mathbf{E}_{\text{DC}}=(0,0,0)$, $\mathbf{E}_{\text{DC}}=(5,0,0)$ and $\mathbf{E}_{\text{DC}}=(10,0,0)$ V/nm. It is worth noting that the chosen maximum (and minimum) strength of the static fields ensures that the electrons bound for $\mathbf{E}_{\text{DC}}=0$ remain bound to the molecule even in the presence of the static field.
An explicit treatment of the electric field is essential to prevent the electrons from travelling to the edges of the box at a large field amplitude. While modern polarisation theory \cite{umari_2002} can address this problem, we retain the explicit treatment of $\mathbf{E}_{\text{DC}}$ as it has proven useful in previous studies to describe the evaporation dynamics of atoms from molecules and periodic systems \cite{feibelman_2001, sanchez_2004, tamura_2012, Silaeva_2015}.\\
\indent
The nucleus-electron interaction is modelled using the pseudopotential method within the local density approximation (LDA) \cite{perdew_1992} for the static DFT electron exchange-correlation and the adiabatic LDA (ALDA) for the TDDFT simulations. In this framework, the Si(OH)$_4$ molecule has 32 valence electrons.

\section{Results and discussion}

\subsection{Experimental results}
\begin{figure}[htb!]
   \centering
\includegraphics[width=0.7\linewidth]{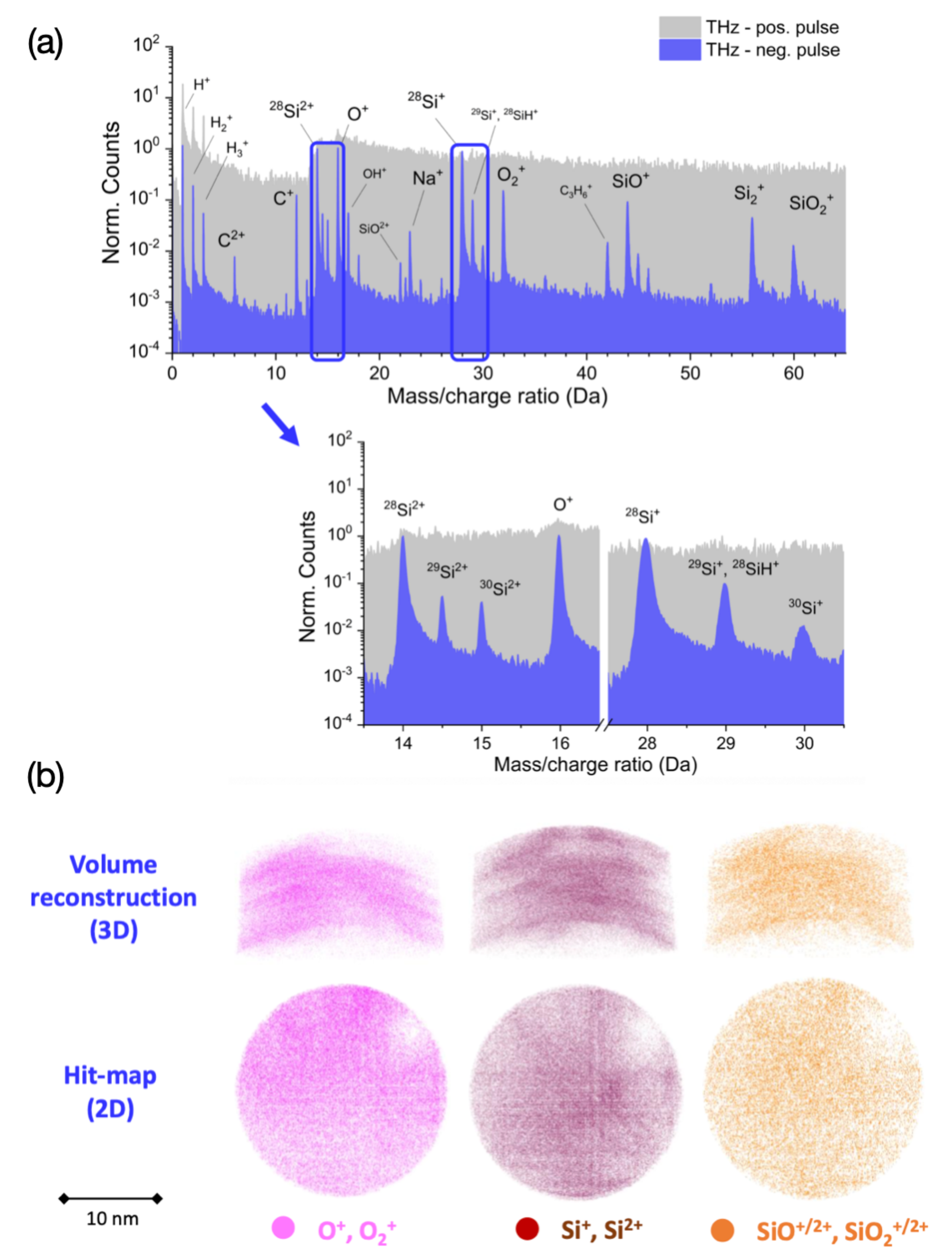}\caption{\label{fig:figure1a} \textbf{(a)} Mass spectrum measured with a THz-assisted atom probe analysis with negative pulses (blue) and positive pulses (grey) of an amorphous silicon dioxide sample. The data sets consist of approximately 4 (1) $\times$10$^5$ ions collected at a bias voltage $U_{\mathrm{bias}}=4 $(9) kV, a THz amplitude of 150 (200) kV/cm, an evaporation rate of 0.08 (0.003) ions/pulse at $T=50$ K, each for negative (positive) pulses. Zoom on the Si$^{2+}$ peak and its isotopes; \textbf{(b)} 2D detection map and 3D atomic distribution of Si (brown dots), oxygen atoms (pink dots) and silicon dioxide (orange dots), reconstructed with a start radius of 30 nm and a cone angle of 6$^{\circ}$, for negative THz pulses.}
\end{figure}
For the analysis of silica samples, the THz amplitude for negative pulses was reduced by approximately 10\% of its maximum and the DC voltage was automatically adjusted to maintain a constant detection rate of 8\% ions per pulse. The mass spectrum in Fig. \ref{fig:figure1a}(a) shows peaks of hydrogen and its molecular ions H$_2^+$ and H$_3^+$. Due to the low repetition rate of our laser system, the sample surface in the high vacuum chamber is contaminated by residual hydrogen, which is field evaporated or field desorbed by the THz pulse. The spectrum also shows silicon ions in the 2+ charge state at 14, 14.5 and 15 Da and in the 1+ state at 28, 29 and 30 Da, carbon ions at 6 and 12 Da for 2+ and 1+ charge states respectively and oxygen at 16 Da. A peak at 23 Da suggests the presence of sodium, probably a residue from the sol-gel preparation. Molecular ions of silica, SiO and SiO$_2$, are observed at 44 and 60 Da for the 1+ charge state and less prominently at 22 and 30 Da for the 2+ state. The peak at 30 Da overlaps with the isotope 30Si$^+$ and affects the natural isotopic distribution for the peaks at 28, 29 and 30 Da. The spectrum also shows the presence of other molecular ions such as OH, OH$_2$ at 42 and 56 Da, C$_3$H$_5^+$ and Si$_2^+$.\\
\indent
For positive THz pulses, the THz amplitude was kept at its maximum and the DC voltage was automatically adjusted to maintain a constant detection rate of 0.3\% ions per pulse. The detection rate is lower than for negative pulses as we cannot increase the DC voltage further due to the increase in static evaporation between the THz pulses.
The mass spectra are shown in Fig. \ref{fig:figure1a}(a) (grey colour).
Only hydrogen peaks (H, H$_2^+$ and H$_3^+$) and a prominent peak from 14 Da are visible.  
The zoom on the spectral range between 13.5 and 17.5 Da reported under Fig. \ref{fig:figure1a}(a) shows that when positive THz pulses are used, all peaks are masked by the strong background noise.\\
\indent
The presence of strong background noise and very large peaks indicates thermally assisted evaporation in the scenario of positive THz pulses. The positive pulses heat up the sample, leading to heating and cooling cycles at the tip of the sample, resulting in the emission of ions and the observed noise and peaks. In contrast, the Si$^{2+}$ and Si$^{+}$ peaks at negative THz pulses exhibit a signal-to-noise ratio (SNR) of more than 10$^3$ and a mass resolution at 10\% of the peak maximum of 146 and 140, respectively (see the zoomed image in Fig. \ref{fig:figure1a}(a)).
These values are comparable to those obtained with UV laser-assisted APT on oxides \cite{marquis2010probing,vella2011field}. The asymmetry of the Si$^{2+}$ and Si$^{+}$ peaks, characterised by a long tail on the right-hand side, indicates that thermal effects also play a role in the evaporation process with negative THz pulses, albeit to a lesser extent.\\
\indent 
\begin{figure*}[htb!]
   \centering
        \includegraphics[width=1\textwidth]{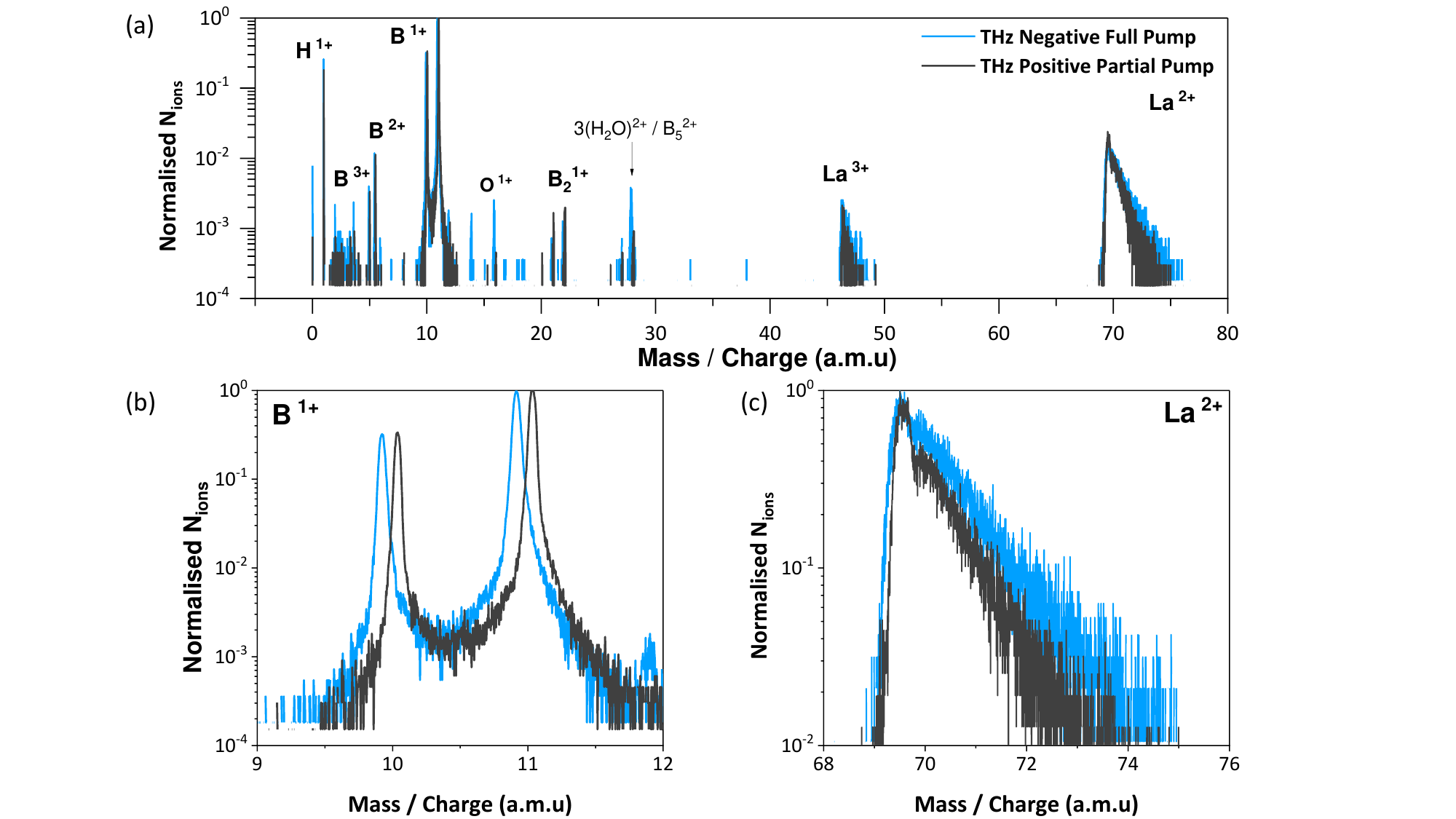} %
    \caption{\label{fig:figure2}\textbf{(a)}
Mass spectrum measured with a THz-assisted APT with negative (blue) and positive (black) pulses of a pure LaB$_6$ sample. The data sets consist of about 10$^5$ ions collected at a bias voltage $U_{\mathrm{bias}}=6.35$ kV, a full THz amplitude of 200 kV/cm for negative pulses and a reduced amplitude of 100 kV/cm for positive pulses, an evaporation rate of 0.04 ions/pulse at $T=50$ K. \textbf{(b)} Zoom of B$^+$ mass peaks using the semi-logarithmic scale. \textbf{(c)} Zoom of the La$^{2+}$ mass peak using the semi-logarithmic scale.} 
\end{figure*}
For the analysis of the LaB$_6$ samples, the THz amplitude was set to its maximum for the negative THz pulse and reduced for the positive THz pulse in order to obtain the same DC voltage and a constant detection rate of 4\% ions/pulse.
The mass spectra are shown in Fig. \ref{fig:figure2}(a) and exhibit similar patterns for both configurations. Zooming in on the B$^+$ peak (Fig. \ref{fig:figure2}(b)) reveals the presence of two boron isotopes and a smaller peak of BH$^+$ molecular ions for negative THz pulses. The peaks width is larger for negative THz pulses, despite a constant signal-to-noise ratio. The shift of the mass/charge position is attributed to the acceleration/deceleration of the ions by the THz pulse \cite{karam2023thz}. Zooming in on the La$^{2+}$ peak shows larger peaks for negative pulses, with two distinct components: a narrow peak associated with fast field evaporation processes, which is higher for positive THz pulses, and a broader peak associated with thermal evaporation processes, which is higher for negative pulses, as discussed in Ref. \cite{karam2023thz, karam2024thz}. The higher power of negative THz pulses leads to higher thermal contributions.\\
\indent 
To better understand the different behaviour of LaB$_6$ and silica samples, we measured the amplification of the THz field by a silica sample using a faster method than in our previous work on LaB$_6$ \cite{karam2023thz}. When a positive voltage of a few volts is applied to the sample, the THz pulses (positive and negative) emit electrons that reach the detector. By increasing the positive voltage applied to the sample, the detection rate of the electrons gradually decreases until it reaches the level of background noise, so that we can measure the voltage that stops the detection of the electrons ($V_{\mathrm{stop}}$). We used numerical simulations to determine the relationship between $V_{\mathrm{stop}}$ and the THz pulse amplitude in kV.
To simulate the path of the electrons in the atom probe chamber, we used the commercial software LORENTZ-2E V10.2 to analyse particle trajectories. The chamber is modelled as an X-axis symmetric structure, where the apex is represented by a cone with a semi-circular tip of 70 nm radius and an angle of 87$^\circ$. The tip is a few millimetres long and is held at a constant DC voltage of several kilovolts, together with a monocyclic THz pulse. The microchannel plate (MCP) detector is shown as a grounded flat plate with a height of 4 cm, and the hemispherical shield with a radius of 15 cm is also grounded. In the experimental setup, the distance between the tip and the detector is set to $L = 9$ cm (Fig. \ref{fig:Geometric_Field}(a)).

\begin{figure}[h]
    \centering
    \includegraphics[width=\linewidth]{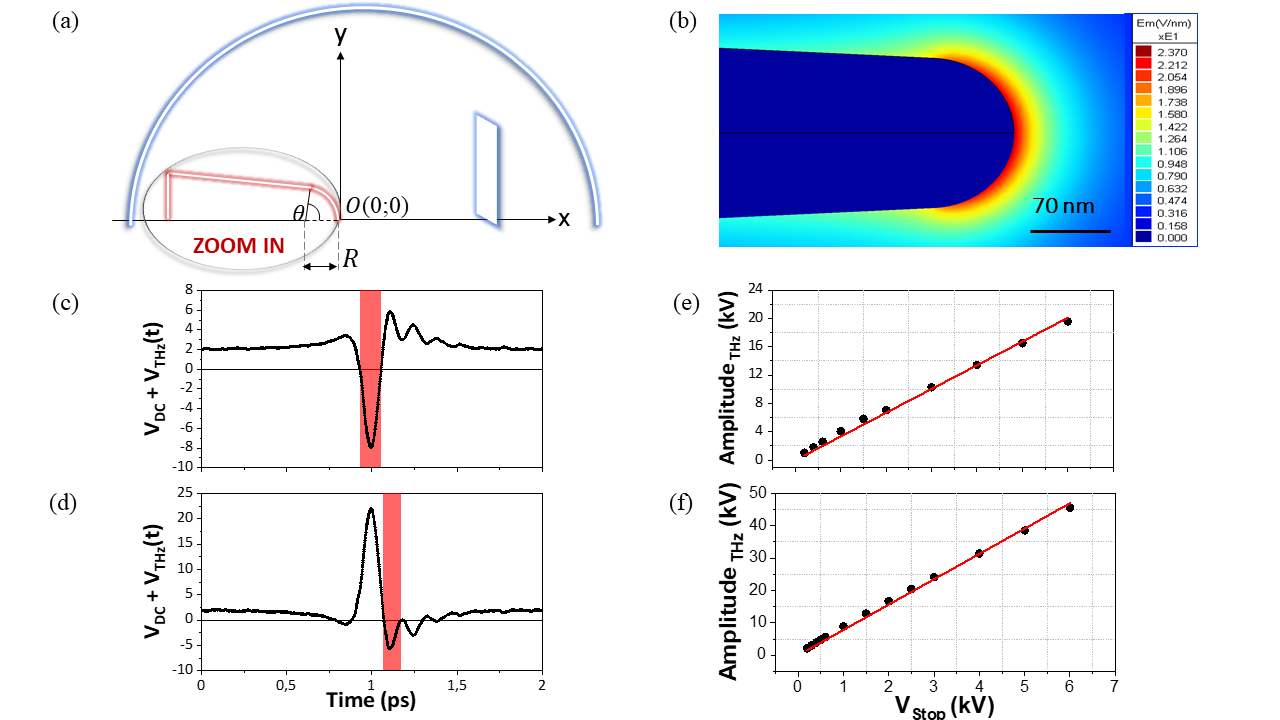}  
\caption{(a) Schematic representation of the APT chamber: the areas in blue are grounded, in red at the combined potential of $V_{\mathrm{DC}}$ and $V_{\mathrm{THz}}(t)$. The tip is shown by a zoom in. (b) Example of an electric field distribution for $V_{\mathrm{DC}}$ = 7 kV and a THz pulse amplitude of 4.5 kV. (c) Negative THz pulse with an amplitude of 10 kV at a DC voltage of 2 kV, and (d) positive THz pulse with an amplitude of 18 kV at a DC voltage of 2 kV. Simulation results for THz amplitude as a function of the blocking voltage for (e) negative and (f) positive THz pulses, respectively. The red lines correspond to the linear fit with the equation: $\mathrm{THz}_{\mathrm{amplitude}}=3.3 \cdot V_{\mathrm{stop}}$ and $\mathrm{THz}_{\mathrm{amplitude}}=7.8 \cdot V_{\mathrm{stop}}$ respectively.}
    \label{fig:Geometric_Field}
\end{figure}

The distribution of the electric field around the tip was calculated using the Boundary Element Method (BEM) in a quasi-transient mode to treat time-dependent electric field phenomena \cite{BEM}. The problem was treated as a series of stationary problems (Fig. \ref{fig:Geometric_Field}(b)). The trajectories of the electrons were calculated by discretizing the applied THz pulse into hundreds of points, each representing a specific departure time. This approach allows a detailed temporal segmentation of the electron emission process. The total potential applied to the sample consists of a continuous background voltage $V_{\mathrm{DC}}$ and a time-varying pulse amplitude $V_{\mathrm{THz}}(t)$. The emission position and time were specified and the electron trajectories were calculated using a fifth-order adaptive Runge-Kutta algorithm (RK5) to solve the dynamic equation $m_e a = neF$, where $m_e$ is the electron mass, $a$ the acceleration, $n$ the electron density and $F$ the total field.
Electron emission was specified within the time interval in which the potential was below zero (highlighted by the red box in Fig. \ref{fig:Geometric_Field}(c-d)). To determine the relationship between the THz pulse amplitude and $V_{\mathrm{stop}}$, we started with a fixed $V_{\mathrm{DC}}$ of 2 kV and a high amplitude of 10 kV. At this amplitude and voltage, all emitted electrons reached the detector. We then gradually reduced the pulse amplitude until no more electrons were detected. This process was repeated for different values of $V_{\mathrm{DC}}$.
Fig. \ref{fig:Geometric_Field}(e-f) shows the variation of the THz amplitude as a function of the blocking static potential in the case of a negative or positive THz pulse. In both cases, a linear dependence with a slope of 3.3 and 7.8 is observed.\\
\indent 
For silica samples, we measured $V_{\mathrm{stop}}$ = 2.25 kV for negative pulses and 0.4 kV for positive pulses, corresponding to THz amplitudes of 7.4 kV and 3.1 kV, respectively. By analysing the ratio of Si$^{2+}$/Si$^+$ ions using the Kingham curves, we calculated the evaporation field $F_{\mathrm{evap}}$ to be 20 V/nm. This field is achieved with a DC voltage of 7.5 kV. By applying the relationship $F_{\mathrm{evap}} = 7.5/(\beta R)$, we have determined the geometric factor $\beta R$, where $R$ is the radius of the sample and $\beta$ is a factor dependent on the tip geometry. The THz field can then be calculated as $F_{\mathrm{THz}} = A_{\mathrm{THz}}/(\beta R)$ = 20 V/nm for negative pulses and 8 V/nm for positive pulses, where $A_{\mathrm{THz}}$ is the THz pulse amplitude. Considering that the THz field before interaction with the sample is 15 MV/m, we observed an amplification factor of 1300 for negative pulses, similar to that measured for LaB$_6$ samples. For positive pulses, however, the amplification decreased by a factor of 2.5.

\subsection{Theoretical analysis: Orthosilic acid molecule}

To explain these puzzling experimental results between samples with different electronic conductivity, we decided to perform a thorough theoretical analysis of the evaporation process of atoms in an APT setup.\\
\indent 
Silica matrices are amorphous materials with different bonding schemes and substitutional defects. SiOH$_4$ is the basic component of such matrices and its size allows affordable first-principles calculations.
In our framework, atomic ions and electrons are exposed to tunable external fields that are either static ($\mathrm{E_{DC}}$
), time-dependent ($\mathrm{E_{laser}}$) or both. The total external field (static + time-dependent) is expressed by \cite{Silaeva_2015}:

\begin{equation}\label{eq:ext_fields}
v_{\mathrm{ext}}(\mathbf{x},t) = \mathbf{x} \cdot \hat{\mathbf{e}} \left[E_{\mathrm{DC}} + E_{\mathrm{laser}} \ \text{exp}\left( -\frac{(t-t_0)^2}{2\sigma^2} \right) \text{sin}(\omega t) \right]
\end{equation}

\noindent where $\hat{\mathbf{e}}$ is the polarisation vector, $\frac{\omega}{2 \pi}= 5$ THz, $t_0$=50 fs (corresponding to the maximum amplitude of the laser pulse), and $\sigma$=14.13 fs, $E_{\mathrm{DC}}$ is always positive while $E_{\mathrm{laser}}$ is either positive (i.e. with the same sign of the static field) or negative.

The interaction of the laser field with the molecule is described using the dipole approximation. In our simulations, the box edge is $a=2$ nm and the laser wavelength is $\lambda \approx 6 \times 10^4$ nm, which fulfils the dipole condition $\lambda \gg a$. All calculations use the OCTOPUS code \cite{octopus_2015,octopus_2020}, in which we have implemented an additional internal function to handle static external fields interacting with ions.

\begin{figure}
	\includegraphics[scale=0.47]{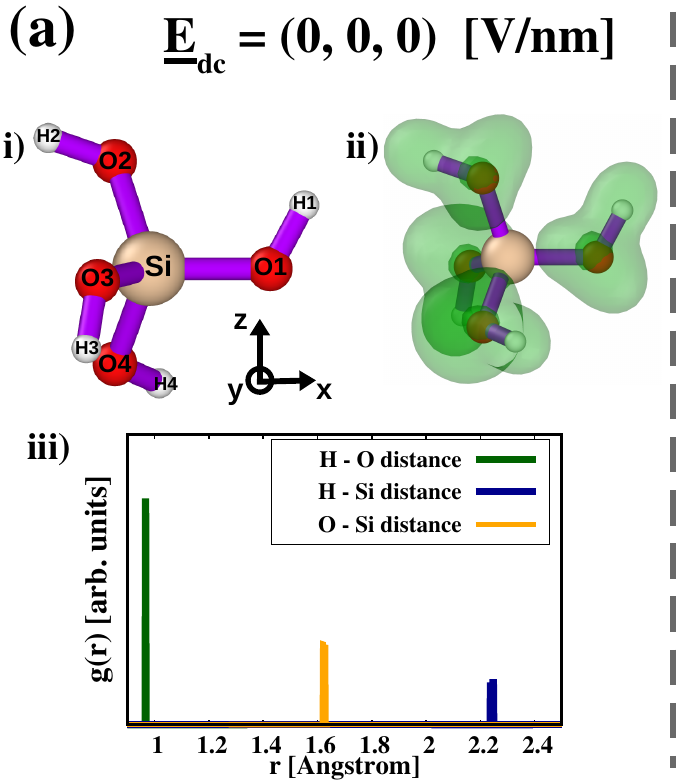}
 	\includegraphics[scale=0.47]{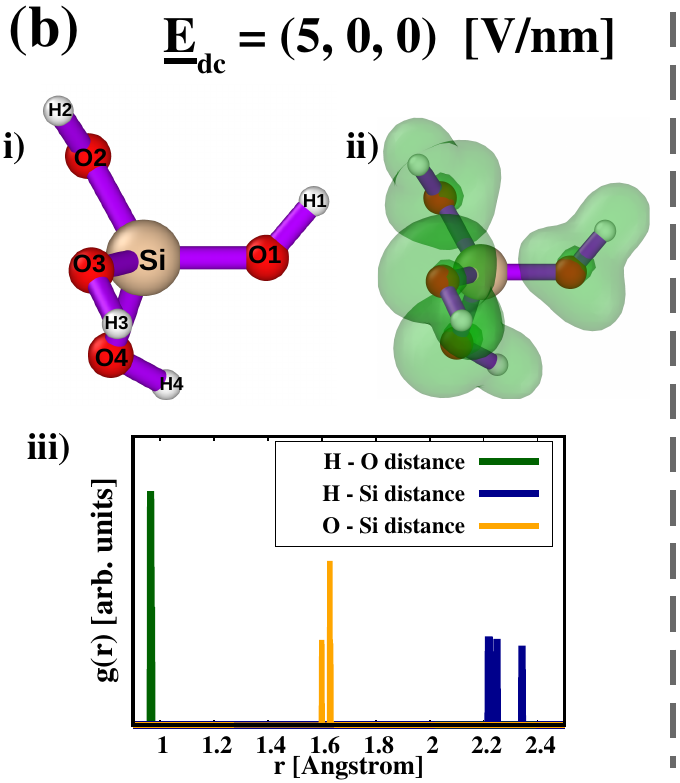}
	\includegraphics[scale=0.47]{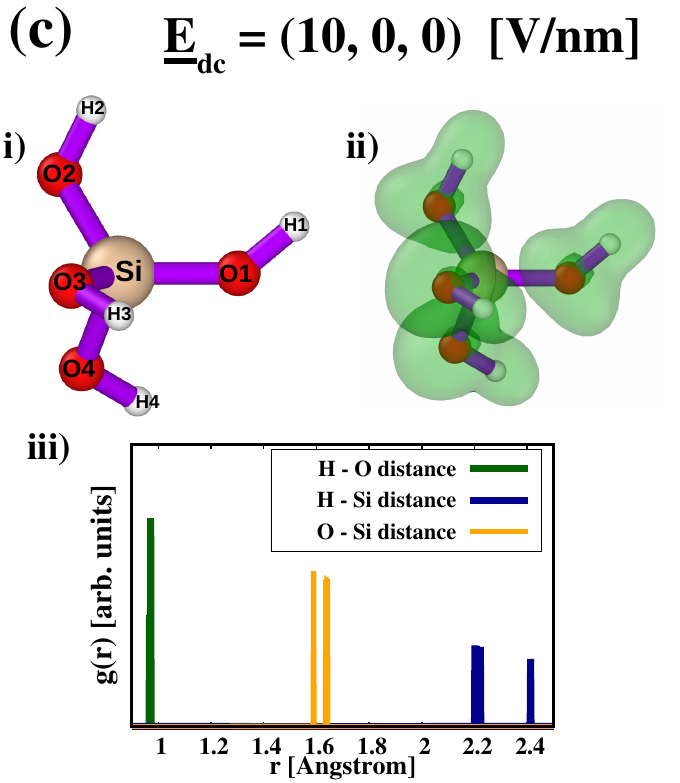}  
 \caption{Si(OH)$_4$ molecule as a function of the static external field: (a) $\mathbf{E}_{\mathrm DC}=(0,0,0)$ V/nm; (b) $\mathbf{E}_{\mathrm DC}=(5,0,0)$ V/nm; (c) $\mathbf{E}_{\mathrm DC}=(10,0,0)$ V/nm. The insets (i) at the top left of panels (a), (b) and (c) are the geometries, while in the insets at the top right (ii) the electron localisation function (ELF) at the isosurface level 0.66 has been added to the structure; the insets at the bottom (iii) show the corresponding pair distribution functions. The geometries were optimised by clamping the Si atom, the $y$ and $z$ coordinates of the oxygen atom O$1$ and the $y$ coordinate of the corresponding hydrogen atom H$1$.}
    \label{fig:SiOH4_geom}
\end{figure}

\begin{figure}[hbt!]
\centering
\includegraphics[scale=0.82]{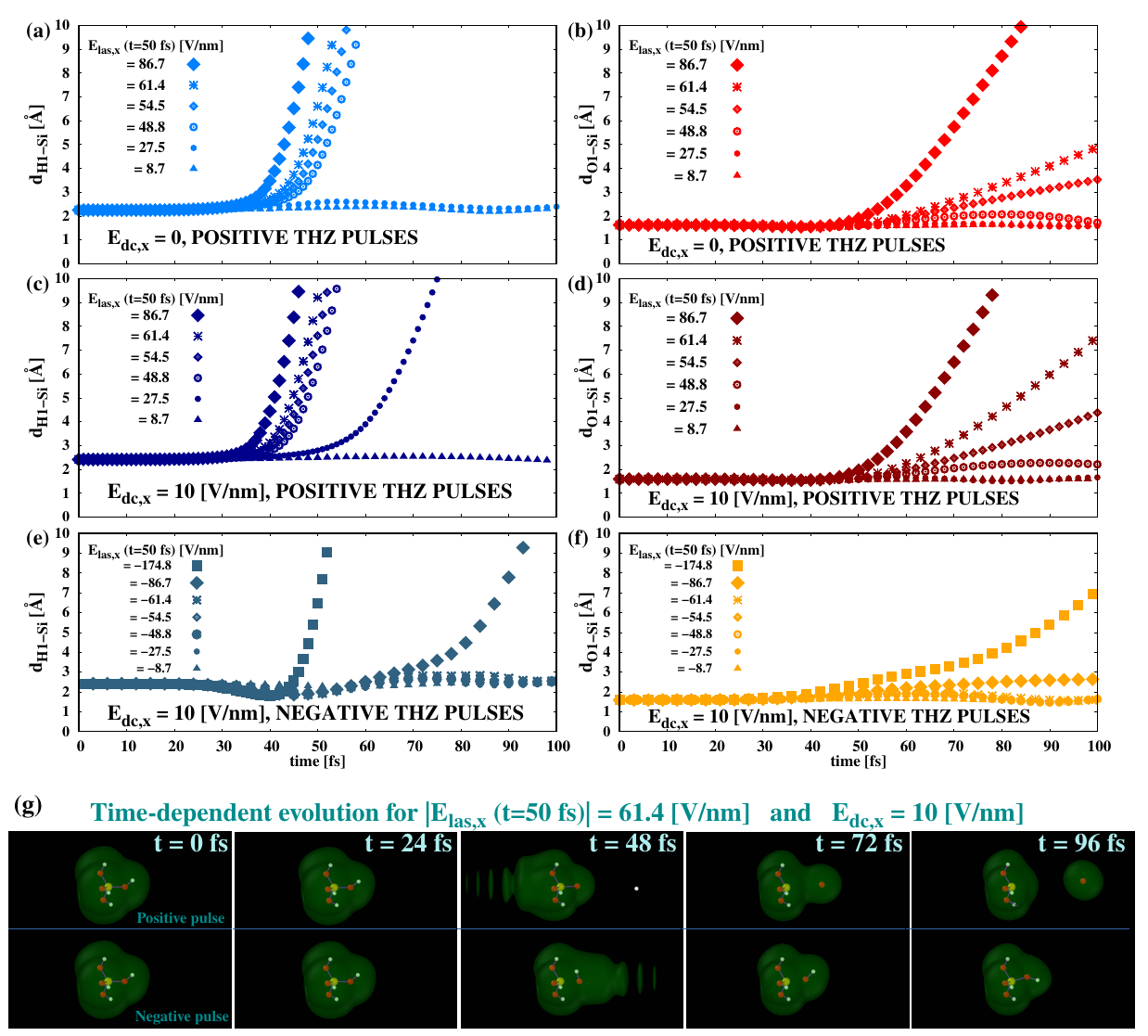} 
\caption{(a,c,e): Distances between the hydrogen atom H1 and silicon as a function of time. (b,d,f): Distances between the oxygen atom O1 and silicon as a function of time. The labels follow the convention of Figure \ref{fig:SiOH4_geom}. (a,b,c,d) correspond to the distances due to positive THz pulses, while (e,f) represent negative pulses. (g) Plot of the isosurface of electronic density equal to $10^{-4}$ and ions for five different snapshots of the temporal evolution for $\mathrm{E}_{{\mathrm{laser}},x}(t=50~ \mathrm{fs})=61.4$ V/nm; the upper plots show the evolution with the positive THz pulse, the lower ones with the negative pulse.}
    \label{fig:distance_thz}
\end{figure}

The DFT equilibrium structure of the Si(OH)$_4$ molecule at each value of the external static field is used as the initial configuration for the temporal evolution of the system. The molecule has a tetrahedrical shape with the silicon atom coordinated by four hydroxyl groups \cite{belton_2012}. The molecule's geometry for $\mathbf{E}_{\mathrm DC}=(0,0,0)$ [V/nm] is shown in Fig. \ref{fig:SiOH4_geom}(a). The pair distribution function in the lower part of the figure shows that at equilibrium the bond lengths are the same for all O-H and O-Si pairs.
When an external static electric field ($\mathbf{E}_{\text{DC}}$) is applied, the molecule polarises in the direction of the field, as shown in Fig. \ref{fig:SiOH4_geom}(b) and Fig. \ref{fig:SiOH4_geom}(c) for the fields $\mathbf{E}_{\text{DC}}=(5,0,0)$ and $\mathbf{E}_{\text{DC}}=(10,0,0)$ V/nm, respectively.
The more intense the static field is, the more strongly the hydroxyl groups are aligned in the direction of the field, which increases the $\widehat{\mathrm{Si-O1-H1}}$ angle in the Si(OH)$_4$ molecule compared to the unpolarised case (Fig. (\ref{fig:SiOH4_geom}a)) and also the bond distance between Si and H$1$ to 2.4 \AA ~(see the labelling of the atoms in the insets i) of Fig. \ref{fig:SiOH4_geom}(a),(b),(c)). \\
\indent During the dynamics, the positions of the silicon atom and the O2H2, O3H3 and O4H4 ions are fixed to prevent the molecule from drifting. We are particularly interested in the evaporation of the O1 and H1 atoms in response to static and oscillating fields, which is similar to the evaporation of single atoms from a solid surface \cite{Silaeva_2015}. 
In the time-dependent simulations, we focussed on the extreme values of the static field: $\mathrm{E}_{{\mathrm{DC}},x}=0$ and $\mathrm{E}_{{\mathrm{DC}},x}=10$ V/nm (see Fig. \ref{fig:distance_thz}(a),(b) and (c),(d),(e),(f))), where the geometries in Fig. \ref{fig:SiOH4_geom}(a),(c) are used as the initial configuration. We note that these configurations are only a small subset of the many possible alignments between the external fields and the molecule. We have specifically chosen the configuration where the O1-Si bond is aligned along the $x$-axis to uniquely evaluate the contributions of the Coulomb force, the static electric field, and the time-dependent laser for each value of the external fields. These results should therefore be regarded as an upper bound for the real physical system, as in reality there may be different angles between fields and bonds as well as other influencing factors such as defects.

\indent
We find a clear difference in the behaviour of the system depending on the intensity of the external fields and the sign of the THz pulses. The results of the time-dependent simulations are shown in Fig. \ref{fig:distance_thz}. In particular, the distance between the hydrogen atom H$1$ and the Si atom as a function of time is shown in Figs. \ref{fig:distance_thz}(a),(c),(e) and the distance between the oxygen atom O1 and the Si atom is shown in Figs. \ref{fig:distance_thz}(b),(d),(f). Different symbols stand for different laser field intensities and different panels refer to different $E_{\mathrm{DC}}$ fields; the Figs. \ref{fig:distance_thz}(a),(b),(c),(d) show the behaviour at positive THz pulses, while Figs. \ref{fig:distance_thz}(e),(f) under the negative ones.\\
\indent
A positive THz field with a peak amplitude of 27.5 V/nm can evaporate hydrogen in the presence of a static external field of $E_{{\mathrm{DC}},x}=10$ V/nm (see blue dots in Fig. \ref{fig:distance_thz}(c)), but not without the static field (Fig. \ref{fig:distance_thz}(a)). We stress that the prediction of this behaviour on a static level by calculating the potential energy surface alone is not correct in principle due to the non-linear interplay between the fields that lead to different ionisations of the system. For oxygen, a minimum amplitude of 54.5 V/nm is required to break the Si-O1 bond, as indicated by the red diamonds in Figs. \ref{fig:distance_thz}(b),(d). 
If one uses negative THz pulses instead, the evaporation thresholds are significantly larger (see Figs. \ref{fig:distance_thz}(e), (f)). In particular, we note that the absolute value of the peak of the time-dependent field for evaporation with negative pulses is about three times as large as the peak in the positive case. The trend of the absolute values of the critical fields is surprisingly in the same direction as our experimental results for LaB$_6$ and not for silica, despite a different behaviour between positive and negative pulses. Indeed, for LaB$_6$
the evaporation starts with a negative pulse three times larger than the threshold of the positive pulse.
This can be explained by the fact that the opposite signs of the static and time-dependent fields make the total field smaller than in the case where both fields are aligned in the same direction and have the same sign. As we will show below, this reduces the ionisation of the system, which is the main mechanism by which surface bonds are broken and atoms evaporate.\\
\indent
When atoms evaporate, the distance between hydrogen and silicon increases rapidly for both positive and negative THz pulses, which facilitates the determination of the evaporation point. However, the slower increase in the distance between oxygen and silicon makes it more difficult to determine this point within the time frame of our simulations. Nevertheless, Ehrenfest dynamics provides valuable information about the forces and velocities acting on the atoms and allows us to calculate the kinetic energy of the oxygen atom. When analysing this data together with the charge density, one can recognise the important role of ionisation in the evaporation process. Intense laser fields lead to increased ionisation of the fixed part of the molecule. The positively charged evaporating H1 and O1 atoms experience a stronger Coulomb repulsion from the fixed part of the molecule, pushing them further away, even if no static external fields are present. Conversely, if a negative THz pulse is used, the partial cancellation of the effect of the positive static field leads to lower ionisation, resulting in a higher threshold for evaporation even at high intensities.
\begin{table}
    \centering
    \begin{tabular}{cc||ccc|cc}
         $E_{\text{dc},x}$ $\left[\frac{V}{\text{nm}}\right]$ & $E_{\text{las},x}(t=50 \ fs)$ $\left[\frac{V}{\text{nm}}\right]$  & $N^*_{mol}$  & $N^*_{H1}$ & $N^*_{O{1}}$ & $d_{H{1}-Si}$ [\AA] & $d_{O1-Si}$ [\AA] \\
         \hline
         \hline
         0 & 8.7 & - & - & - & 2.30 & 1.61 \\
         10 & 8.7 & 0.1 & - & - & 2.36 & 1.59 \\
         10 & -8.7 & - & - & - & 2.48 & 1.59 \\ 
         \hline
         0 & 27.5 & 0.8  & - & - & 2.39 & 1.58 \\
         10 & 27.5 & 0.6 & 1.0 & - & $>$ 10 & 1.58 \\
        10 & -27.5 & 0.7 & - & - & 2.41 & 1.65 \\ 
         \hline
         0 & 41.1 & 1.2  & 1.0 & - &  $>$ 10 & 1.60 \\
         10 & 41.1 & 2.0 & 1.0 & - &  $>$ 10 & 1.49  \\
         10 & -41.1 & 1.1 & - & - & 2.60 & 1.64 \\
         \hline
         0 & 48.8 & 1.6 & 1.0 & -  & $>$ 10  & 1.71 \\
        10 & 48.8 & 2.6  & 1.0 & - & $>$ 10 & 2.20  \\
         10 & -48.8 & 2.0 & - & - & 2.55 & 1.64 \\
         \hline
         0 & 54.5 & 2.4  & 1.0 & 0.5 & $>$ 10 & 3.52   \\
         10 & 54.5 & 2.5 & 1.0 & 0.9 & $>$ 10 & 4.37  \\
         10 & -54.5 & 2.4 & - & -  & 2.58 & 1.62 \\         
         \hline         
         0 & 61.4 & 2.7 & 1.0 & 0.7 & $>$ 10  & 4.88 \\
         10 & 61.4 & 2.8  & 1.0 & 1.0 & $>$ 10 & 6.27 \\
         10 & -61.4 & 1.9 & - & - & 2.55 & 1.59    \\    
         \hline
         0 & 86.7 & 3.9 & 1.0 & 1.9 & $>$ 10 & $>$ 10 \\
         10 & 86.7 & 4.0 & 1.0 & 2.1 & $>$ 10 & $>$ 10 \\
         10 & -86.7 & 4.2 & 1.0 & - & $>$ 10 & 2.64 \\ 
         \hline
         10 & -174.8 & 9.8 & 1.0 & 1.5 & $>$ 10 & 7.11 \\ 
    \hline     
    \end{tabular}
    \caption{Data from the Ehrenfest dynamics simulations: $d_{\mathrm{H1-Si}}$ and $d_{\mathrm{O1-Si}}$ represent the distances between the hydrogen atom H$1$ and the Si atom, and between the oxygen atom O$1$ and the Si atom, respectively. $N^*_{\mathrm{mol}}$ denotes the final total charge of the fixed molecule, while $N^*_{\mathrm{O1}}$ and $N^*_{\mathrm{H1}}$ indicate the charges of the evaporating oxygen and hydrogen atoms. The latter two values are only given if the distances between these atoms and the Si exceed 3 \AA. The number of electrons emitted is determined by the sum $N^*_{\mathrm{mol}}+N^*_{\mathrm{O1}}+N^*_{\mathrm{H1}}$. In the table, the symbol ``-'' indicates the neutral state.}
    \label{tab:results}
\end{table}
Fig. \ref{fig:distance_thz}(g) shows the electronic density at the isosurface level of $10^{-4}$ and the ions at different time steps during the simulation for an absolute value of the THz pulse peak amplitude of 61.4 V/nm, in both positive (upper panel) and negative (lower panel) cases. A significant difference in the charge densities polarized in opposite directions is observed near the peak of the laser ($t\sim 48$ fs).
The results for the charge state and the distance reached by the atoms at the end of the simulation are summarized in Table \ref{tab:results}. The hydrogen atom always evaporates fully ionized with a charge of +1, while the charge of the evaporated oxygen atom increases with the laser amplitude when using a positive THz pulse. The total charges induced by the negative pulse are consistently larger due to the lower ionization.\\
\indent
The theoretical prediction that positive THz pulses have a lower threshold amplitude than negative pulses to induce atomic evaporation in silica is not supported by our experimental analysis of the mass spectra. Our results show that the critical amplitudes of the THz fields required for the evaporation of H1 and O1 are close to the experimental values \cite{vella_2021} although, for insulating systems with large band gaps, only negative pulses are experimentally efficient for the evaporation process.
The discrepancy between theory and experiments can be explained by the fact that a single Si(OH)$_4$ molecule in a vacuum is not an accurate model for a needle-shaped amorphous silica, which more closely resembles a bulk system. In a bulk system, excited electrons have a finite lifetime and can be extracted by interaction with electromagnetic fields if the energy of the external field exceeds either the band gap plus the work function (for insulators) or only the work function (for metals), provided they are all at the bottom of the conduction band.
The subsequent behavior of the electrons crucially depends on whether the material is an insulator or a metallic system. 

\subsection{Discussion}

To explain the effect of electron mobility and finite lifetime on the evaporation of atoms from the surface of a bulk material under static and time-dependent fields and to reconcile the experimental and theoretical results, which seem to contradict each other with respect to the role of polarity, we have developed a 
model that can explain the outcomes of experimental measurements.\\
\indent
In fact, experimentally we have measured that silica samples amplify the THz field by a factor of 1300 for negative pulses, as for LaB$_6$ samples. For positive pulses, however, the amplification decreased by a factor of 2.5 only for silica samples. Due to the poor electrical conductivity of those samples, the interaction with THz pulses depends on the sign of the pulse. The 5-micron silica tip is attached to a metallic pre-tip made of tungsten and forms a Schottky junction, which only amplifies the THz pulse if the sign of the electric field matches the forward bias of the Schottky diode, i.e. the negative sign.
In the case of the LaB$_6$ samples, the junction is non-existent because LaB$_6$ behaves like a semimetal with excellent electrical conductivity, similar to a metal.\\
\indent
To describe this behavior we have developed a fictitious model of a bulk material that can be used to describe a needle-shaped sample in an APT setup. In this context, we have used the simplest bound system, an H$_2$ molecule, in 1D to keep the simulations as easy as possible. In particular, we have introduced an additional potential barrier that can be tuned to mimic the behaviour of a bulk system, acting either as an obstacle to electron flow in insulating systems (presence of a steep potential barrier) or as a conductive channel in metallic systems (no potential barrier).
\begin{figure}[hbt!]
    \centering
    \includegraphics[width=0.8\textwidth]{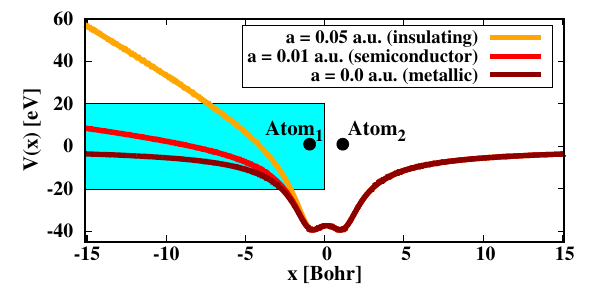}
    \caption{The one-dimensional potential of our toy model described by Eq. \ref{eq:1dpot} is experienced by the two fictitious electrons within the molecule and varies with different left barriers. In this scenario, Atom$_2$ is bound to an atom (Atom$_1$) on the solid surface, symbolised by the cyan rectangle. The steepness of the barrier is influenced by the mobility of the electrons, which depends on either the Fermi level (for metals) or the band gap (for insulators/semiconductors) of the system.}
    \label{fig:pot}
\end{figure}
This configuration is shown in Fig. \ref{fig:pot}. We emphasise that this model aims to interpret the physical mechanisms and does not claim to be realistic in terms of the absolute values of the fields.\\
\indent 
We consider the evaporation of the rightmost atom (Atom$_2$ in Fig. \ref{fig:pot}) from the surface of a solid, where the surface is ideally represented by the Atom$_1$, to which Atom$_2$ is bound, and a potential barrier that reflects the nature of the material, whether it is a metal, a semiconductor or an insulator. As said before, we assume that Atom$_1$ and Atom$_2$ are hydrogen-like and each consist of an electron and a proton.\\
\indent 
Similar to the case of Si(OH)$_4$, evaporation is triggered by a static potential, which is always positive, and a THz pulse (Eq. \ref{eq:ext_fields}). 
In the Supplementary Materials, we also show the response of the system to multi-cycle UV fields (Fig. S.1) and to single-cycle IR fields (Figs. S.2 and S.3). In atomic units, the potential felt by an electron in the 1D system is given by
\begin{equation}\label{eq:1dpot}
    V(x) = a\cdot \left(\frac{\pi}{2}-\text{atan}\left(10^4\cdot(x-X_1)\right)\right)|x-X_1|-\frac{1}{\sqrt{1+(x-X_1)^2}}-\frac{1}{\sqrt{1+(x-X_2)^2}}, 
\end{equation}
where the parameter $a$ controls the barrier that makes the system more metal-like ($a \to 0$) or more insulator-like ($a > 0$); $X_1$ and $X_2$ are the positions of the two atoms (see Fig. \ref{fig:pot}). To simplify the simulations, we use a modified Coulomb potential $1/\sqrt{1+x^2}$ with the same core-level energy as the hydrogen atom (i.e. -0.5 atomic units) instead of a pure Coulomb potential $1/x$. We also stress that the barrier is linear. If a sufficiently strong static electric field is applied, its effect can in principle be cancelled out. This mimics the behaviour of insulating systems in which the electrons can overcome the band gap and populate the conduction bands by applying a large electric field.

\begin{figure}[hbt!]
\centering
\includegraphics[width=0.49\textwidth]{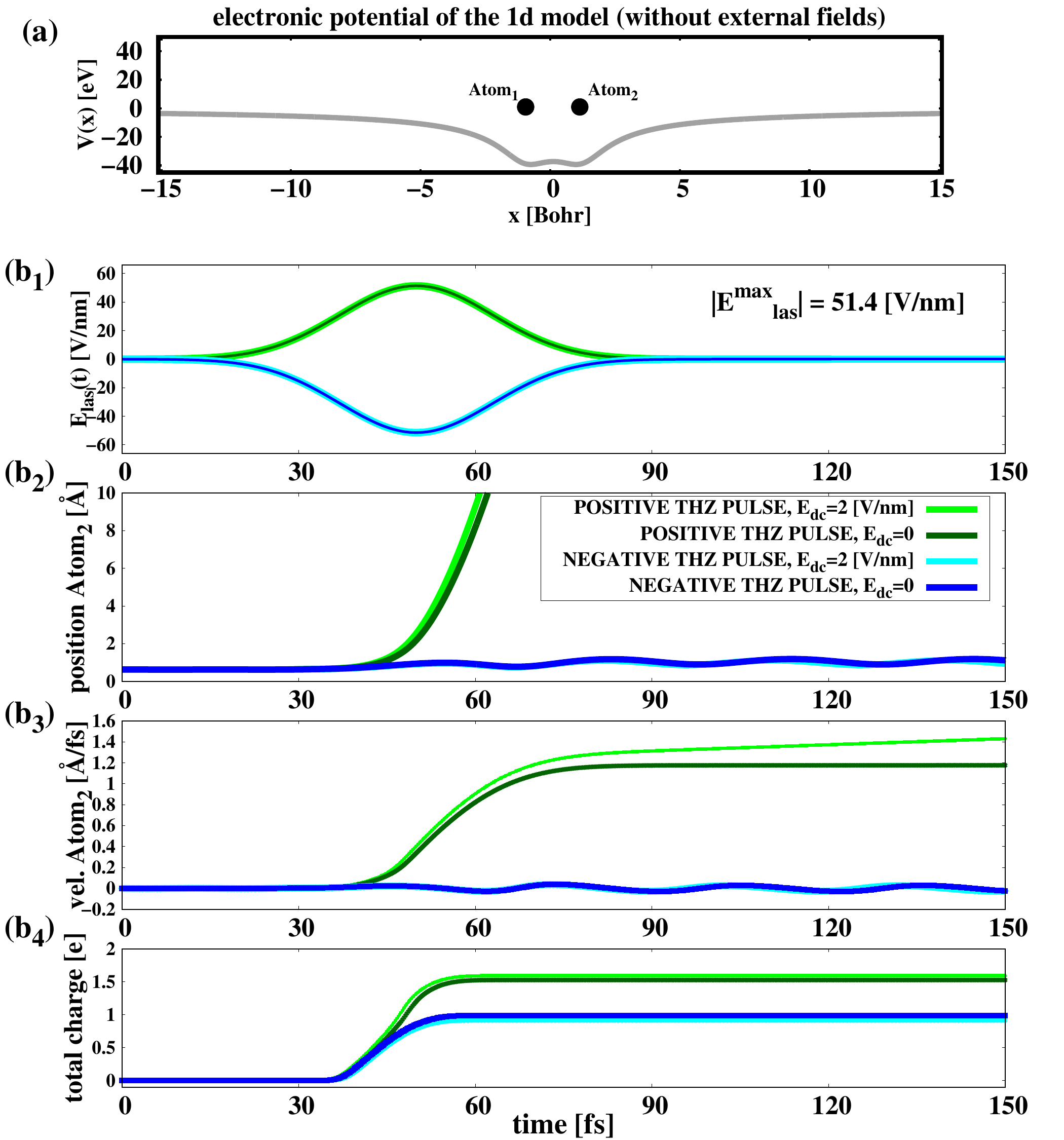}
\includegraphics[width=0.49\textwidth]{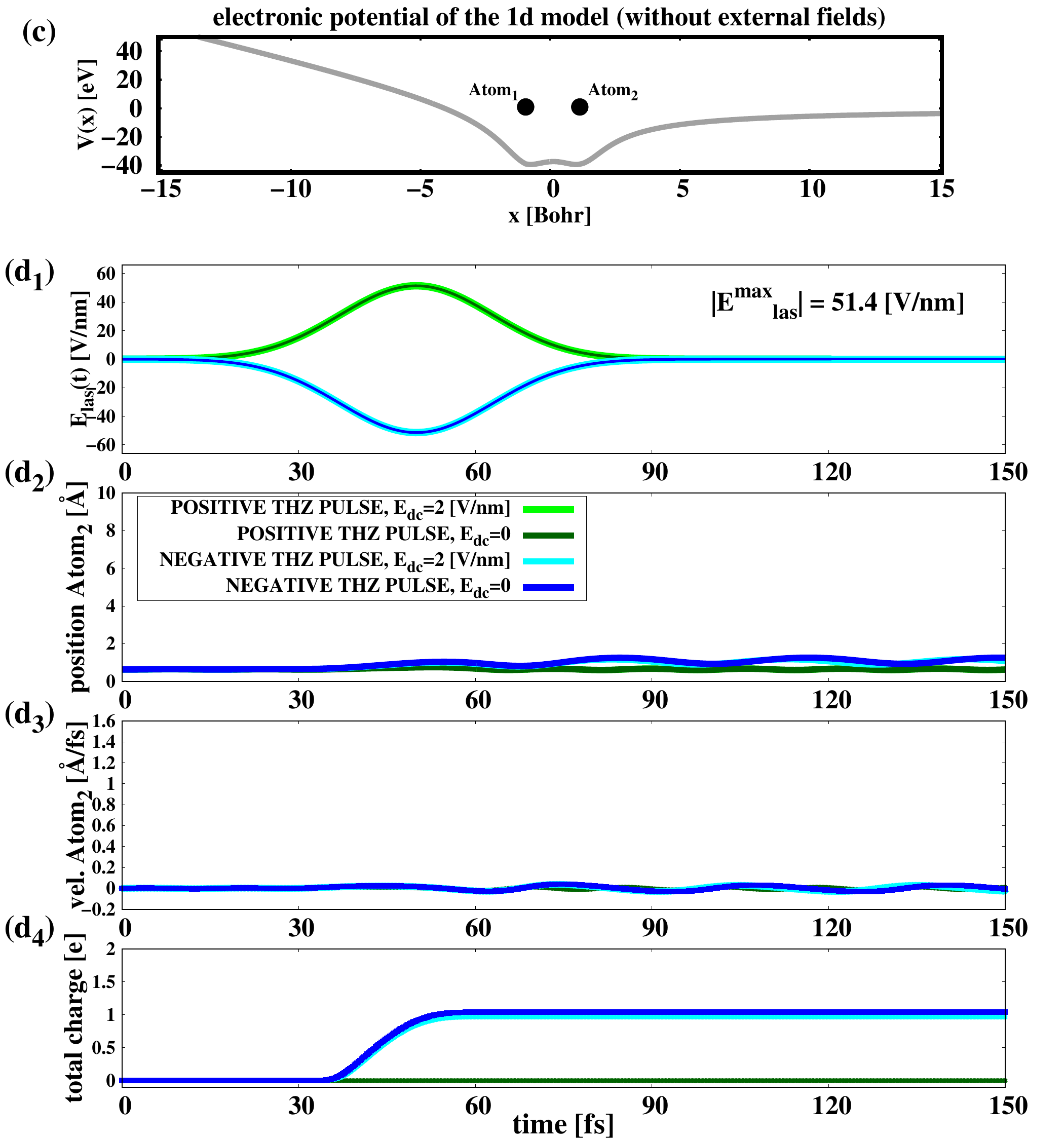}
\caption{Comparison of the results obtained with the two potentials in (a) and (c) for a fixed amplitude of the THz pulse in (b$_1$) and (d$_1$). The maximum amplitude is 0.1 atomic units. The light and dark green curves correspond to the positive pulse, while the cyan and blue lines represent the negative pulse. In (b$_2$) and (d$_2$) the position of Atom$_2$ is shown as a function of time and in (b$_3$) and (d$_3$) its velocity in the different cases. Finally, the total charge inside the box is given in (b$_4$) and (d$_4$).
\label{fig:metalinsulator0.1}}
\end{figure}

\begin{figure}[hbt!]\label{fig:tm012}
\centering
\includegraphics[width=0.48\textwidth]{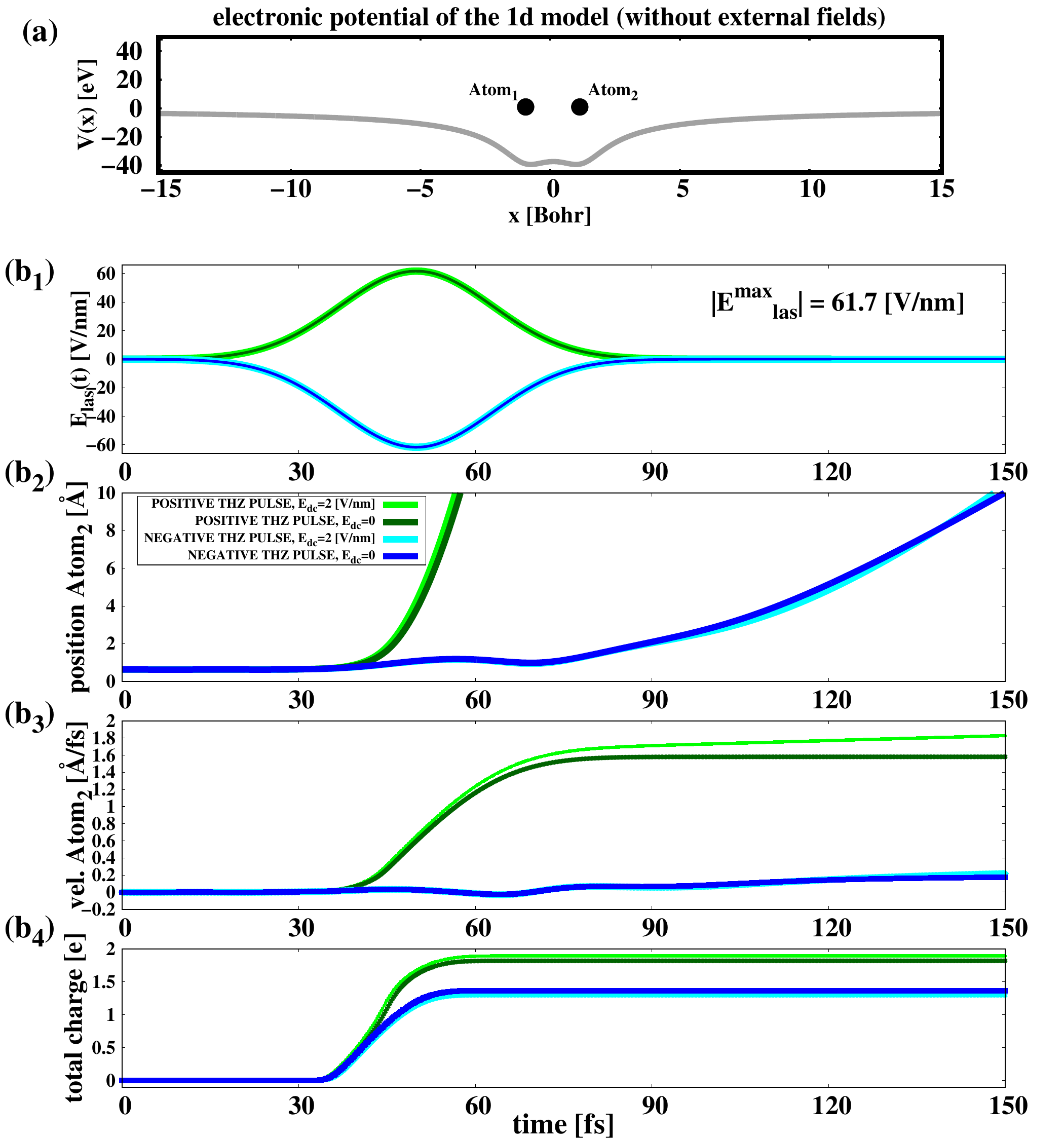}
\includegraphics[width=0.48\textwidth]{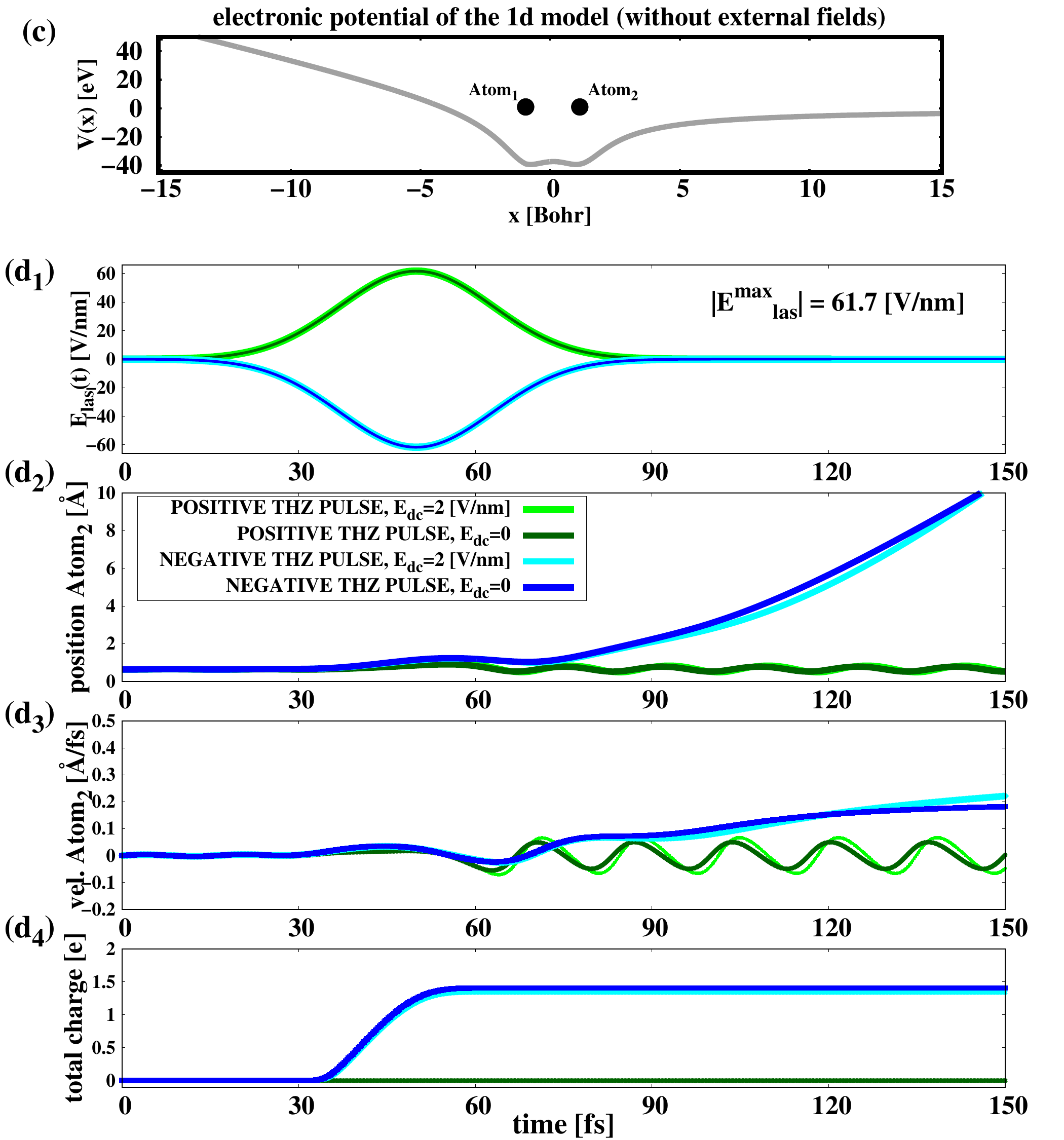}
\caption{Comparison of the results obtained with the two potentials in (a) and (c) for a fixed amplitude of the THz pulse in (b$_1$) and (d$_1$). The maximum amplitude is 0.12 atomic units.
The light and dark green curves correspond to the positive pulse, while the cyan and blue lines represent the negative pulse. In (b$_2$) and (d$_2$) the position of Atom$_2$ is shown as a function of time and in (b$_3$) and (d$_3$) its velocity in the different cases. Finally, (b$_4$) and (d$_4$) show the total charge inside the box.
\label{fig:metalinsulator0.12}}
\end{figure}
The results obtained with this oversimplified model are shown in Figs. \ref{fig:metalinsulator0.1} and \ref{fig:metalinsulator0.12}.
In Fig. \ref{fig:metalinsulator0.1} we compare the behaviour of the molecule without a barrier (a) and with a barrier (c). Without the left potential barrier, a peak field of 51.4 V/nm (see Fig. \ref{fig:metalinsulator0.1}(b$_1$)) allows Atom$_2$ to evaporate under a positive THz pulse (light and dark green curves in Fig. \ref{fig:metalinsulator0.1}(b$_2$)); while under a negative pulse Atom$_2$ only oscillates around its equilibrium position (cyan and blue lines). The velocity of the evaporating atom increases linearly with a static electric field (light green) and remains constant without a field (dark green) (Fig. \ref{fig:metalinsulator0.1}(b$_3$)). In the presence of the DC potential, positive pulses induce strong ionisation, leading to electrostatic repulsion that dominates the system dynamics, while negative pulses do not ionise sufficiently and keep the system bound. The degrees of ionisation vary with different combinations of static and laser field, with the highest degree of ionisation being achieved with a positive THz pulse with a static field (green curve in Fig. \ref{fig:metalinsulator0.1}(b$_4$)).\\
\indent 
However, when a left potential barrier is added, the dynamics under a positive THz field changes significantly. Evaporation no longer takes place (see Fig. \ref{fig:metalinsulator0.1}(d$_2$),(d$_3$)), due to the inability to ionize the system, resulting in a constant total charge. The negative dynamics is less affected with respect to the case without barrier as the electrons are removed from the barrier-free right-hand side. Therefore, the barrier prevents positive THz pulses from releasing electrons, which remain trapped. This behaviour mimics insulating systems in which electrons do not flow freely and their rearrangement under electromagnetic fields leads to polarisation of the sample. Thus, while the amplitude of the pulse is sufficient for evaporation without a barrier, it is not sufficient if the left potential barrier is present. 
\\
\indent 
In Fig. \ref{fig:metalinsulator0.12} we compare the behaviour of the model with (rigth) and without (left) a barrier under a larger amplitude of the time-dependent field compared to Fig. \ref{fig:metalinsulator0.1}.
On the left side of Fig. \ref{fig:metalinsulator0.12} Atom$_2$ evaporates both when a positive THz pulse is used (light green and dark green lines) and when a negative pulse is used (cyan and blue curves). However, the efficiency of the positive pulse compared to the negative pulse is clear. On the right side, evaporation is not observed with positive pulses but only with negative THz (blue and cyan lines). Therefore, the threshold of negative pulses becomes smaller than the one for positive one, in agreement with the experimental findings for silica.

\section{Conclusions}

In this article, we have performed a comprehensive theoretical and experimental analysis to investigate the mechanisms responsible for the evaporation of cations from nanoneedle-shaped tips of materials of different nature.
We used a THz-assisted APT setup to study the mass spectra of in-house grown silica samples, which are insulators, used in biological applications to trap proteins and LaB$_6$ samples, a metallic system. In our setup, a total potential was applied to the samples, which consisted of a continuous background voltage $V_{\mathrm{DC}}$ and a time-varying pulse amplitude $V_{\mathrm{THz}}(t)$ in the THz frequency range. Our main finding is that single-cycle high-intensity THz pulses can evaporate non-conducting materials such as amorphous silica and that the interaction of silica with THz pulses depends on polarity, in contrast to LaB$_6$ where this dependence was not observed. This dependence could also not be found when multi-cycle pulses in a different frequency range such as UV were used, but can theoretically be predicted for single-cycle pulses in a different frequency range such as the near IR, assuming a higher pulse intensity than THz pulses. When comparing the behaviour of LaB$_6$ and silicon dioxide, we observed in the latter case a significant amplification of the THz field exceeding 1300 times, similar to the LaB$_6$ samples, but only for negative pulses. For positive pulses, the amplification decreased by a factor of 2.5. This finding has a profound impact on the interpretation of the evaporation processes.\\
\indent 
We attribute these diversified results to the different degree of ionisation of the samples, with amorphous silica having a more effective ionisation for negative pulses due to its insulating nature. The amplification of THz pulses in insulating materials occurs when the sign of the electric field matches the forward bias of the Schottky diode, resulting in evaporation. In contrast, metals such as LaB$_6$, which have a high electrical conductivity, lack this junction.
In order to understand the evaporation process, the conductivity differences between the samples must therefore be carefully considered.\\ 
\indent
However, our TDDFT simulations, which are crucial for studying realistic samples, did not provide any insights into this process for an orthosilicic acid molecule. This is mainly
due to the fact that a single molecule can hardly reproduce the behaviour of solids, be it semiconductors or insulators, especially in terms of electron mobility. To interpret the experimental measurements, we have therefore developed a simplified model at the vacuum-nanoneedles interface and shown that the dynamics under a positive THz field changes significantly when a potential barrier is introduced to simulate insulator/semiconductor materials, which hinders ionisation. Evaporation stops when a left potential barrier is added as the system cannot be ionised. The behaviour of our model system with and without a barrier under a stronger time-dependent field is able to capture these effects on cation evaporation.\\
The polarity dependence of single-cycle pulses in the emission of cations from non-conducting materials can be extended beyond the THz range investigated here to all spectral ranges in which single-cycle pulses with high intensity can be generated.\\
\indent 
Future work includes the application of our model to more complex structures and the extension of real-time real-space TDDFT simulations to study atomic evaporation under static and laser fields.
This interdisciplinary approach expands the capabilities of APT and holds promise for other THz-based analyses offering new insights into material dynamics in complex biological environments.

\begin{acknowledgments} This action has received funding from the European Union under grant agreement No 10104665
\end{acknowledgments}


\bibliography{biblio}

\providecommand{\noopsort}[1]{}\providecommand{\singleletter}[1]{#1}%
\begin{thebibliography}{48}%
\makeatletter
\providecommand \@ifxundefined [1]{%
 \@ifx{#1\undefined}
}%
\providecommand \@ifnum [1]{%
 \ifnum #1\expandafter \@firstoftwo
 \else \expandafter \@secondoftwo
 \fi
}%
\providecommand \@ifx [1]{%
 \ifx #1\expandafter \@firstoftwo
 \else \expandafter \@secondoftwo
 \fi
}%
\providecommand \natexlab [1]{#1}%
\providecommand \enquote  [1]{``#1''}%
\providecommand \bibnamefont  [1]{#1}%
\providecommand \bibfnamefont [1]{#1}%
\providecommand \citenamefont [1]{#1}%
\providecommand \href@noop [0]{\@secondoftwo}%
\providecommand \href [0]{\begingroup \@sanitize@url \@href}%
\providecommand \@href[1]{\@@startlink{#1}\@@href}%
\providecommand \@@href[1]{\endgroup#1\@@endlink}%
\providecommand \@sanitize@url [0]{\catcode `\\12\catcode `\$12\catcode `\&12\catcode `\#12\catcode `\^12\catcode `\_12\catcode `\%12\relax}%
\providecommand \@@startlink[1]{}%
\providecommand \@@endlink[0]{}%
\providecommand \url  [0]{\begingroup\@sanitize@url \@url }%
\providecommand \@url [1]{\endgroup\@href {#1}{\urlprefix }}%
\providecommand \urlprefix  [0]{URL }%
\providecommand \Eprint [0]{\href }%
\providecommand \doibase [0]{https://doi.org/}%
\providecommand \selectlanguage [0]{\@gobble}%
\providecommand \bibinfo  [0]{\@secondoftwo}%
\providecommand \bibfield  [0]{\@secondoftwo}%
\providecommand \translation [1]{[#1]}%
\providecommand \BibitemOpen [0]{}%
\providecommand \bibitemStop [0]{}%
\providecommand \bibitemNoStop [0]{.\EOS\space}%
\providecommand \EOS [0]{\spacefactor3000\relax}%
\providecommand \BibitemShut  [1]{\csname bibitem#1\endcsname}%
\let\auto@bib@innerbib\@empty
\bibitem [{\citenamefont {Sadeghian}\ and\ \citenamefont {Islam}(2011)}]{sadeghian2011ultralow}%
  \BibitemOpen
  \bibfield  {author} {\bibinfo {author} {\bibfnamefont {R.~B.}\ \bibnamefont {Sadeghian}}\ and\ \bibinfo {author} {\bibfnamefont {M.~S.}\ \bibnamefont {Islam}},\ }\bibfield  {title} {\bibinfo {title} {Ultralow-voltage field-ionization discharge on whiskered silicon nanowires for gas-sensing applications},\ }\href@noop {} {\bibfield  {journal} {\bibinfo  {journal} {Nature materials}\ }\textbf {\bibinfo {volume} {10}},\ \bibinfo {pages} {135} (\bibinfo {year} {2011})}\BibitemShut {NoStop}%
\bibitem [{\citenamefont {Liu}\ and\ \citenamefont {Miller}(2007)}]{liu2007field}%
  \BibitemOpen
  \bibfield  {author} {\bibinfo {author} {\bibfnamefont {J.-F.}\ \bibnamefont {Liu}}\ and\ \bibinfo {author} {\bibfnamefont {G.~P.}\ \bibnamefont {Miller}},\ }\bibfield  {title} {\bibinfo {title} {Field-assisted nanopatterning},\ }\href@noop {} {\bibfield  {journal} {\bibinfo  {journal} {The Journal of Physical Chemistry C}\ }\textbf {\bibinfo {volume} {111}},\ \bibinfo {pages} {10758} (\bibinfo {year} {2007})}\BibitemShut {NoStop}%
\bibitem [{\citenamefont {Ovchinnikov}\ \emph {et~al.}(2019)\citenamefont {Ovchinnikov}, \citenamefont {Chefonov}, \citenamefont {Mishina},\ and\ \citenamefont {Agranat}}]{ovchinnikov2019second}%
  \BibitemOpen
  \bibfield  {author} {\bibinfo {author} {\bibfnamefont {A.}~\bibnamefont {Ovchinnikov}}, \bibinfo {author} {\bibfnamefont {O.}~\bibnamefont {Chefonov}}, \bibinfo {author} {\bibfnamefont {E.}~\bibnamefont {Mishina}},\ and\ \bibinfo {author} {\bibfnamefont {M.}~\bibnamefont {Agranat}},\ }\bibfield  {title} {\bibinfo {title} {Second harmonic generation in the bulk of silicon induced by an electric field of a high power terahertz pulse},\ }\href@noop {} {\bibfield  {journal} {\bibinfo  {journal} {Scientific reports}\ }\textbf {\bibinfo {volume} {9}},\ \bibinfo {pages} {1} (\bibinfo {year} {2019})}\BibitemShut {NoStop}%
\bibitem [{\citenamefont {Klarskov}\ \emph {et~al.}(2017)\citenamefont {Klarskov}, \citenamefont {Kim}, \citenamefont {Colvin},\ and\ \citenamefont {Mittleman}}]{klarskov2017nanoscale}%
  \BibitemOpen
  \bibfield  {author} {\bibinfo {author} {\bibfnamefont {P.}~\bibnamefont {Klarskov}}, \bibinfo {author} {\bibfnamefont {H.}~\bibnamefont {Kim}}, \bibinfo {author} {\bibfnamefont {V.~L.}\ \bibnamefont {Colvin}},\ and\ \bibinfo {author} {\bibfnamefont {D.~M.}\ \bibnamefont {Mittleman}},\ }\bibfield  {title} {\bibinfo {title} {Nanoscale laser terahertz emission microscopy},\ }\href@noop {} {\bibfield  {journal} {\bibinfo  {journal} {ACS Photonics}\ }\textbf {\bibinfo {volume} {4}},\ \bibinfo {pages} {2676} (\bibinfo {year} {2017})}\BibitemShut {NoStop}%
\bibitem [{\citenamefont {Dombi}\ \emph {et~al.}(2020)\citenamefont {Dombi}, \citenamefont {P{\'a}pa}, \citenamefont {Vogelsang}, \citenamefont {Yalunin}, \citenamefont {Sivis}, \citenamefont {Herink}, \citenamefont {Sch{\"a}fer}, \citenamefont {Gro{\ss}}, \citenamefont {Ropers},\ and\ \citenamefont {Lienau}}]{dombi2020strong}%
  \BibitemOpen
  \bibfield  {author} {\bibinfo {author} {\bibfnamefont {P.}~\bibnamefont {Dombi}}, \bibinfo {author} {\bibfnamefont {Z.}~\bibnamefont {P{\'a}pa}}, \bibinfo {author} {\bibfnamefont {J.}~\bibnamefont {Vogelsang}}, \bibinfo {author} {\bibfnamefont {S.~V.}\ \bibnamefont {Yalunin}}, \bibinfo {author} {\bibfnamefont {M.}~\bibnamefont {Sivis}}, \bibinfo {author} {\bibfnamefont {G.}~\bibnamefont {Herink}}, \bibinfo {author} {\bibfnamefont {S.}~\bibnamefont {Sch{\"a}fer}}, \bibinfo {author} {\bibfnamefont {P.}~\bibnamefont {Gro{\ss}}}, \bibinfo {author} {\bibfnamefont {C.}~\bibnamefont {Ropers}},\ and\ \bibinfo {author} {\bibfnamefont {C.}~\bibnamefont {Lienau}},\ }\bibfield  {title} {\bibinfo {title} {Strong-field nano-optics},\ }\href@noop {} {\bibfield  {journal} {\bibinfo  {journal} {Reviews of Modern Physics}\ }\textbf {\bibinfo {volume} {92}},\ \bibinfo {pages} {025003} (\bibinfo {year} {2020})}\BibitemShut {NoStop}%
\bibitem [{\citenamefont {LaRue}\ \emph {et~al.}(2015)\citenamefont {LaRue}, \citenamefont {Katayama}, \citenamefont {Lindenberg}, \citenamefont {Fisher}, \citenamefont {{\"O}str{\"o}m}, \citenamefont {Nilsson},\ and\ \citenamefont {Ogasawara}}]{larue2015thz}%
  \BibitemOpen
  \bibfield  {author} {\bibinfo {author} {\bibfnamefont {J.~L.}\ \bibnamefont {LaRue}}, \bibinfo {author} {\bibfnamefont {T.}~\bibnamefont {Katayama}}, \bibinfo {author} {\bibfnamefont {A.}~\bibnamefont {Lindenberg}}, \bibinfo {author} {\bibfnamefont {A.~S.}\ \bibnamefont {Fisher}}, \bibinfo {author} {\bibfnamefont {H.}~\bibnamefont {{\"O}str{\"o}m}}, \bibinfo {author} {\bibfnamefont {A.}~\bibnamefont {Nilsson}},\ and\ \bibinfo {author} {\bibfnamefont {H.}~\bibnamefont {Ogasawara}},\ }\bibfield  {title} {\bibinfo {title} {Thz-pulse-induced selective catalytic co oxidation on ru},\ }\href@noop {} {\bibfield  {journal} {\bibinfo  {journal} {Physical review letters}\ }\textbf {\bibinfo {volume} {115}},\ \bibinfo {pages} {036103} (\bibinfo {year} {2015})}\BibitemShut {NoStop}%
\bibitem [{\citenamefont {Silaeva}\ \emph {et~al.}(2014)\citenamefont {Silaeva}, \citenamefont {Arnoldi}, \citenamefont {Karahka}, \citenamefont {Deconihout}, \citenamefont {Menand}, \citenamefont {Kreuzer},\ and\ \citenamefont {Vella}}]{silaeva2014dielectric}%
  \BibitemOpen
  \bibfield  {author} {\bibinfo {author} {\bibfnamefont {E.}~\bibnamefont {Silaeva}}, \bibinfo {author} {\bibfnamefont {L.}~\bibnamefont {Arnoldi}}, \bibinfo {author} {\bibfnamefont {M.}~\bibnamefont {Karahka}}, \bibinfo {author} {\bibfnamefont {B.}~\bibnamefont {Deconihout}}, \bibinfo {author} {\bibfnamefont {A.}~\bibnamefont {Menand}}, \bibinfo {author} {\bibfnamefont {H.}~\bibnamefont {Kreuzer}},\ and\ \bibinfo {author} {\bibfnamefont {A.}~\bibnamefont {Vella}},\ }\bibfield  {title} {\bibinfo {title} {Do dielectric nanostructures turn metallic in high-electric dc fields?},\ }\href@noop {} {\bibfield  {journal} {\bibinfo  {journal} {Nano letters}\ }\textbf {\bibinfo {volume} {14}},\ \bibinfo {pages} {6066} (\bibinfo {year} {2014})}\BibitemShut {NoStop}%
\bibitem [{\citenamefont {Schultze}\ \emph {et~al.}(2013)\citenamefont {Schultze}, \citenamefont {Bothschafter}, \citenamefont {Sommer}, \citenamefont {Holzner}, \citenamefont {Schweinberger}, \citenamefont {Fiess}, \citenamefont {Hofstetter}, \citenamefont {Kienberger}, \citenamefont {Apalkov}, \citenamefont {Yakovlev} \emph {et~al.}}]{schultze2013controlling}%
  \BibitemOpen
  \bibfield  {author} {\bibinfo {author} {\bibfnamefont {M.}~\bibnamefont {Schultze}}, \bibinfo {author} {\bibfnamefont {E.~M.}\ \bibnamefont {Bothschafter}}, \bibinfo {author} {\bibfnamefont {A.}~\bibnamefont {Sommer}}, \bibinfo {author} {\bibfnamefont {S.}~\bibnamefont {Holzner}}, \bibinfo {author} {\bibfnamefont {W.}~\bibnamefont {Schweinberger}}, \bibinfo {author} {\bibfnamefont {M.}~\bibnamefont {Fiess}}, \bibinfo {author} {\bibfnamefont {M.}~\bibnamefont {Hofstetter}}, \bibinfo {author} {\bibfnamefont {R.}~\bibnamefont {Kienberger}}, \bibinfo {author} {\bibfnamefont {V.}~\bibnamefont {Apalkov}}, \bibinfo {author} {\bibfnamefont {V.~S.}\ \bibnamefont {Yakovlev}}, \emph {et~al.},\ }\bibfield  {title} {\bibinfo {title} {Controlling dielectrics with the electric field of light},\ }\href@noop {} {\bibfield  {journal} {\bibinfo  {journal} {Nature}\ }\textbf {\bibinfo {volume} {493}},\ \bibinfo {pages} {75} (\bibinfo {year} {2013})}\BibitemShut {NoStop}%
\bibitem [{\citenamefont {Herink}\ \emph {et~al.}(2012)\citenamefont {Herink}, \citenamefont {Solli}, \citenamefont {Gulde},\ and\ \citenamefont {Ropers}}]{herink2012field}%
  \BibitemOpen
  \bibfield  {author} {\bibinfo {author} {\bibfnamefont {G.}~\bibnamefont {Herink}}, \bibinfo {author} {\bibfnamefont {D.~R.}\ \bibnamefont {Solli}}, \bibinfo {author} {\bibfnamefont {M.}~\bibnamefont {Gulde}},\ and\ \bibinfo {author} {\bibfnamefont {C.}~\bibnamefont {Ropers}},\ }\bibfield  {title} {\bibinfo {title} {Field-driven photoemission from nanostructures quenches the quiver motion},\ }\href@noop {} {\bibfield  {journal} {\bibinfo  {journal} {Nature}\ }\textbf {\bibinfo {volume} {483}},\ \bibinfo {pages} {190} (\bibinfo {year} {2012})}\BibitemShut {NoStop}%
\bibitem [{\citenamefont {Dai}\ \emph {et~al.}(2009)\citenamefont {Dai}, \citenamefont {Karpowicz},\ and\ \citenamefont {Zhang}}]{dai2009coherent}%
  \BibitemOpen
  \bibfield  {author} {\bibinfo {author} {\bibfnamefont {J.}~\bibnamefont {Dai}}, \bibinfo {author} {\bibfnamefont {N.}~\bibnamefont {Karpowicz}},\ and\ \bibinfo {author} {\bibfnamefont {X.-C.}\ \bibnamefont {Zhang}},\ }\bibfield  {title} {\bibinfo {title} {Coherent polarization control of terahertz waves generated from two-color laser-induced gas plasma},\ }\href@noop {} {\bibfield  {journal} {\bibinfo  {journal} {Physical Review Letters}\ }\textbf {\bibinfo {volume} {103}},\ \bibinfo {pages} {023001} (\bibinfo {year} {2009})}\BibitemShut {NoStop}%
\bibitem [{\citenamefont {Sato}\ \emph {et~al.}(2013)\citenamefont {Sato}, \citenamefont {Higuchi}, \citenamefont {Kanda}, \citenamefont {Konishi}, \citenamefont {Yoshioka}, \citenamefont {Suzuki}, \citenamefont {Misawa},\ and\ \citenamefont {Kuwata-Gonokami}}]{sato2013terahertz}%
  \BibitemOpen
  \bibfield  {author} {\bibinfo {author} {\bibfnamefont {M.}~\bibnamefont {Sato}}, \bibinfo {author} {\bibfnamefont {T.}~\bibnamefont {Higuchi}}, \bibinfo {author} {\bibfnamefont {N.}~\bibnamefont {Kanda}}, \bibinfo {author} {\bibfnamefont {K.}~\bibnamefont {Konishi}}, \bibinfo {author} {\bibfnamefont {K.}~\bibnamefont {Yoshioka}}, \bibinfo {author} {\bibfnamefont {T.}~\bibnamefont {Suzuki}}, \bibinfo {author} {\bibfnamefont {K.}~\bibnamefont {Misawa}},\ and\ \bibinfo {author} {\bibfnamefont {M.}~\bibnamefont {Kuwata-Gonokami}},\ }\bibfield  {title} {\bibinfo {title} {Terahertz polarization pulse shaping with arbitrary field control},\ }\href@noop {} {\bibfield  {journal} {\bibinfo  {journal} {Nature Photonics}\ }\textbf {\bibinfo {volume} {7}},\ \bibinfo {pages} {724} (\bibinfo {year} {2013})}\BibitemShut {NoStop}%
\bibitem [{\citenamefont {D{\'e}chard}\ \emph {et~al.}(2018)\citenamefont {D{\'e}chard}, \citenamefont {Debayle}, \citenamefont {Davoine}, \citenamefont {Gremillet},\ and\ \citenamefont {Berg{\'e}}}]{dechard2018terahertz}%
  \BibitemOpen
  \bibfield  {author} {\bibinfo {author} {\bibfnamefont {J.}~\bibnamefont {D{\'e}chard}}, \bibinfo {author} {\bibfnamefont {A.}~\bibnamefont {Debayle}}, \bibinfo {author} {\bibfnamefont {X.}~\bibnamefont {Davoine}}, \bibinfo {author} {\bibfnamefont {L.}~\bibnamefont {Gremillet}},\ and\ \bibinfo {author} {\bibfnamefont {L.}~\bibnamefont {Berg{\'e}}},\ }\bibfield  {title} {\bibinfo {title} {Terahertz pulse generation in underdense relativistic plasmas: From photoionization-induced radiation to coherent transition radiation},\ }\href@noop {} {\bibfield  {journal} {\bibinfo  {journal} {Physical review letters}\ }\textbf {\bibinfo {volume} {120}},\ \bibinfo {pages} {144801} (\bibinfo {year} {2018})}\BibitemShut {NoStop}%
\bibitem [{\citenamefont {Koulouklidis}\ \emph {et~al.}(2020)\citenamefont {Koulouklidis}, \citenamefont {Gollner}, \citenamefont {Shumakova}, \citenamefont {Fedorov}, \citenamefont {Pug{\v{z}}lys}, \citenamefont {Baltu{\v{s}}ka},\ and\ \citenamefont {Tzortzakis}}]{koulouklidis2020observation}%
  \BibitemOpen
  \bibfield  {author} {\bibinfo {author} {\bibfnamefont {A.~D.}\ \bibnamefont {Koulouklidis}}, \bibinfo {author} {\bibfnamefont {C.}~\bibnamefont {Gollner}}, \bibinfo {author} {\bibfnamefont {V.}~\bibnamefont {Shumakova}}, \bibinfo {author} {\bibfnamefont {V.~Y.}\ \bibnamefont {Fedorov}}, \bibinfo {author} {\bibfnamefont {A.}~\bibnamefont {Pug{\v{z}}lys}}, \bibinfo {author} {\bibfnamefont {A.}~\bibnamefont {Baltu{\v{s}}ka}},\ and\ \bibinfo {author} {\bibfnamefont {S.}~\bibnamefont {Tzortzakis}},\ }\bibfield  {title} {\bibinfo {title} {Observation of extremely efficient terahertz generation from mid-infrared two-color laser filaments},\ }\href@noop {} {\bibfield  {journal} {\bibinfo  {journal} {Nature Communications}\ }\textbf {\bibinfo {volume} {11}},\ \bibinfo {pages} {1} (\bibinfo {year} {2020})}\BibitemShut {NoStop}%
\bibitem [{\citenamefont {Wimmer}\ \emph {et~al.}(2014)\citenamefont {Wimmer}, \citenamefont {Herink}, \citenamefont {Solli}, \citenamefont {Yalunin}, \citenamefont {Echternkamp},\ and\ \citenamefont {Ropers}}]{wimmer2014terahertz}%
  \BibitemOpen
  \bibfield  {author} {\bibinfo {author} {\bibfnamefont {L.}~\bibnamefont {Wimmer}}, \bibinfo {author} {\bibfnamefont {G.}~\bibnamefont {Herink}}, \bibinfo {author} {\bibfnamefont {D.~R.}\ \bibnamefont {Solli}}, \bibinfo {author} {\bibfnamefont {S.~V.}\ \bibnamefont {Yalunin}}, \bibinfo {author} {\bibfnamefont {K.}~\bibnamefont {Echternkamp}},\ and\ \bibinfo {author} {\bibfnamefont {C.}~\bibnamefont {Ropers}},\ }\bibfield  {title} {\bibinfo {title} {Terahertz control of nanotip photoemission},\ }\href@noop {} {\bibfield  {journal} {\bibinfo  {journal} {Nature Physics}\ }\textbf {\bibinfo {volume} {10}},\ \bibinfo {pages} {432} (\bibinfo {year} {2014})}\BibitemShut {NoStop}%
\bibitem [{\citenamefont {Li}\ and\ \citenamefont {Jones}(2016)}]{li2016high}%
  \BibitemOpen
  \bibfield  {author} {\bibinfo {author} {\bibfnamefont {S.}~\bibnamefont {Li}}\ and\ \bibinfo {author} {\bibfnamefont {R.}~\bibnamefont {Jones}},\ }\bibfield  {title} {\bibinfo {title} {High-energy electron emission from metallic nano-tips driven by intense single-cycle terahertz pulses},\ }\href@noop {} {\bibfield  {journal} {\bibinfo  {journal} {Nature communications}\ }\textbf {\bibinfo {volume} {7}},\ \bibinfo {pages} {1} (\bibinfo {year} {2016})}\BibitemShut {NoStop}%
\bibitem [{\citenamefont {Vella}\ \emph {et~al.}(2021{\natexlab{a}})\citenamefont {Vella}, \citenamefont {Houard}, \citenamefont {Arnoldi}, \citenamefont {Tang}, \citenamefont {Boudant}, \citenamefont {Ayoub}, \citenamefont {Normand}, \citenamefont {Da~Costa},\ and\ \citenamefont {Hideur}}]{vella2021high}%
  \BibitemOpen
  \bibfield  {author} {\bibinfo {author} {\bibfnamefont {A.}~\bibnamefont {Vella}}, \bibinfo {author} {\bibfnamefont {J.}~\bibnamefont {Houard}}, \bibinfo {author} {\bibfnamefont {L.}~\bibnamefont {Arnoldi}}, \bibinfo {author} {\bibfnamefont {M.}~\bibnamefont {Tang}}, \bibinfo {author} {\bibfnamefont {M.}~\bibnamefont {Boudant}}, \bibinfo {author} {\bibfnamefont {A.}~\bibnamefont {Ayoub}}, \bibinfo {author} {\bibfnamefont {A.}~\bibnamefont {Normand}}, \bibinfo {author} {\bibfnamefont {G.}~\bibnamefont {Da~Costa}},\ and\ \bibinfo {author} {\bibfnamefont {A.}~\bibnamefont {Hideur}},\ }\bibfield  {title} {\bibinfo {title} {High-resolution terahertz-driven atom probe tomography},\ }\href@noop {} {\bibfield  {journal} {\bibinfo  {journal} {Science Advances}\ }\textbf {\bibinfo {volume} {7}},\ \bibinfo {pages} {eabd7259} (\bibinfo {year} {2021}{\natexlab{a}})}\BibitemShut {NoStop}%
\bibitem [{\citenamefont {Karam}\ \emph {et~al.}(2023)\citenamefont {Karam}, \citenamefont {Houard}, \citenamefont {Damarla}, \citenamefont {Rousseau}, \citenamefont {Bhorade},\ and\ \citenamefont {Vella}}]{karam2023thz}%
  \BibitemOpen
  \bibfield  {author} {\bibinfo {author} {\bibfnamefont {M.}~\bibnamefont {Karam}}, \bibinfo {author} {\bibfnamefont {J.}~\bibnamefont {Houard}}, \bibinfo {author} {\bibfnamefont {G.}~\bibnamefont {Damarla}}, \bibinfo {author} {\bibfnamefont {L.}~\bibnamefont {Rousseau}}, \bibinfo {author} {\bibfnamefont {O.}~\bibnamefont {Bhorade}},\ and\ \bibinfo {author} {\bibfnamefont {A.}~\bibnamefont {Vella}},\ }\bibfield  {title} {\bibinfo {title} {Thz driven field emission: energy and time-of-flight spectra of ions},\ }\href@noop {} {\bibfield  {journal} {\bibinfo  {journal} {New Journal of Physics}\ }\textbf {\bibinfo {volume} {25}},\ \bibinfo {pages} {113017} (\bibinfo {year} {2023})}\BibitemShut {NoStop}%
\bibitem [{\citenamefont {Jelic}\ \emph {et~al.}(2017)\citenamefont {Jelic}, \citenamefont {Iwaszczuk}, \citenamefont {Nguyen}, \citenamefont {Rathje}, \citenamefont {Hornig}, \citenamefont {Sharum}, \citenamefont {Hoffman}, \citenamefont {Freeman},\ and\ \citenamefont {Hegmann}}]{jelic2017ultrafast}%
  \BibitemOpen
  \bibfield  {author} {\bibinfo {author} {\bibfnamefont {V.}~\bibnamefont {Jelic}}, \bibinfo {author} {\bibfnamefont {K.}~\bibnamefont {Iwaszczuk}}, \bibinfo {author} {\bibfnamefont {P.~H.}\ \bibnamefont {Nguyen}}, \bibinfo {author} {\bibfnamefont {C.}~\bibnamefont {Rathje}}, \bibinfo {author} {\bibfnamefont {G.~J.}\ \bibnamefont {Hornig}}, \bibinfo {author} {\bibfnamefont {H.~M.}\ \bibnamefont {Sharum}}, \bibinfo {author} {\bibfnamefont {J.~R.}\ \bibnamefont {Hoffman}}, \bibinfo {author} {\bibfnamefont {M.~R.}\ \bibnamefont {Freeman}},\ and\ \bibinfo {author} {\bibfnamefont {F.~A.}\ \bibnamefont {Hegmann}},\ }\bibfield  {title} {\bibinfo {title} {Ultrafast terahertz control of extreme tunnel currents through single atoms on a silicon surface},\ }\href@noop {} {\bibfield  {journal} {\bibinfo  {journal} {Nature Physics}\ }\textbf {\bibinfo {volume} {13}},\ \bibinfo {pages} {591} (\bibinfo {year} {2017})}\BibitemShut {NoStop}%
\bibitem [{\citenamefont {Li}\ and\ \citenamefont {Jones}(2014)}]{li2014ionization}%
  \BibitemOpen
  \bibfield  {author} {\bibinfo {author} {\bibfnamefont {S.}~\bibnamefont {Li}}\ and\ \bibinfo {author} {\bibfnamefont {R.}~\bibnamefont {Jones}},\ }\bibfield  {title} {\bibinfo {title} {Ionization of excited atoms by intense single-cycle thz pulses},\ }\href@noop {} {\bibfield  {journal} {\bibinfo  {journal} {Physical review letters}\ }\textbf {\bibinfo {volume} {112}},\ \bibinfo {pages} {143006} (\bibinfo {year} {2014})}\BibitemShut {NoStop}%
\bibitem [{\citenamefont {Huzayyin}\ \emph {et~al.}(2014)\citenamefont {Huzayyin}, \citenamefont {Chang}, \citenamefont {Lian},\ and\ \citenamefont {Dawson}}]{huzayyin2014interaction}%
  \BibitemOpen
  \bibfield  {author} {\bibinfo {author} {\bibfnamefont {A.}~\bibnamefont {Huzayyin}}, \bibinfo {author} {\bibfnamefont {J.~H.}\ \bibnamefont {Chang}}, \bibinfo {author} {\bibfnamefont {K.}~\bibnamefont {Lian}},\ and\ \bibinfo {author} {\bibfnamefont {F.}~\bibnamefont {Dawson}},\ }\bibfield  {title} {\bibinfo {title} {Interaction of water molecule with au (111) and au (110) surfaces under the influence of an external electric field},\ }\href@noop {} {\bibfield  {journal} {\bibinfo  {journal} {The Journal of Physical Chemistry C}\ }\textbf {\bibinfo {volume} {118}},\ \bibinfo {pages} {3459} (\bibinfo {year} {2014})}\BibitemShut {NoStop}%
\bibitem [{\citenamefont {Karam}\ \emph {et~al.}(2024)\citenamefont {Karam}, \citenamefont {Houard}, \citenamefont {Bhorade}, \citenamefont {Blum},\ and\ \citenamefont {Vella}}]{karam2024thz}%
  \BibitemOpen
  \bibfield  {author} {\bibinfo {author} {\bibfnamefont {M.}~\bibnamefont {Karam}}, \bibinfo {author} {\bibfnamefont {J.}~\bibnamefont {Houard}}, \bibinfo {author} {\bibfnamefont {O.}~\bibnamefont {Bhorade}}, \bibinfo {author} {\bibfnamefont {I.}~\bibnamefont {Blum}},\ and\ \bibinfo {author} {\bibfnamefont {A.}~\bibnamefont {Vella}},\ }\bibfield  {title} {\bibinfo {title} {Thz vs nir laser-assisted atom probe tomography of lab6 samples},\ }\href@noop {} {\bibfield  {journal} {\bibinfo  {journal} {APL Materials}\ }\textbf {\bibinfo {volume} {12}} (\bibinfo {year} {2024})}\BibitemShut {NoStop}%
\bibitem [{\citenamefont {Ullrich}(2012)}]{ullrich}%
  \BibitemOpen
  \bibfield  {author} {\bibinfo {author} {\bibfnamefont {C.~A.}\ \bibnamefont {Ullrich}},\ }\href@noop {} {\emph {\bibinfo {title} {Time-Dependent Density-Functional Theory: Concepts and Applications}}}\ (\bibinfo  {publisher} {Oxford University Press},\ \bibinfo {year} {2012})\BibitemShut {NoStop}%
\bibitem [{\citenamefont {Marx}\ and\ \citenamefont {Hutter}(2009)}]{marx_2009}%
  \BibitemOpen
  \bibfield  {author} {\bibinfo {author} {\bibfnamefont {D.}~\bibnamefont {Marx}}\ and\ \bibinfo {author} {\bibfnamefont {J.}~\bibnamefont {Hutter}},\ }\bibinfo {title} {Getting started: unifying md and electronic structure},\ in\ \href@noop {} {\emph {\bibinfo {booktitle} {Ab Initio Molecular Dynamics: Basic Theory and Advanced Methods}}}\ (\bibinfo  {publisher} {Cambridge University Press},\ \bibinfo {year} {2009})\ pp.\ \bibinfo {pages} {11--84}\BibitemShut {NoStop}%
\bibitem [{\citenamefont {Silaeva}\ \emph {et~al.}(2015)\citenamefont {Silaeva}, \citenamefont {Uchida}, \citenamefont {Suzuki},\ and\ \citenamefont {Watanabe}}]{Silaeva_2015}%
  \BibitemOpen
  \bibfield  {author} {\bibinfo {author} {\bibfnamefont {E.~P.}\ \bibnamefont {Silaeva}}, \bibinfo {author} {\bibfnamefont {K.}~\bibnamefont {Uchida}}, \bibinfo {author} {\bibfnamefont {Y.}~\bibnamefont {Suzuki}},\ and\ \bibinfo {author} {\bibfnamefont {K.}~\bibnamefont {Watanabe}},\ }\bibfield  {title} {\bibinfo {title} {Energetics and dynamics of laser-assisted field evaporation: Time-dependent density functional theory simulations},\ }\href {https://doi.org/10.1103/PhysRevB.92.155401} {\bibfield  {journal} {\bibinfo  {journal} {Phys. Rev. B}\ }\textbf {\bibinfo {volume} {92}},\ \bibinfo {pages} {155401} (\bibinfo {year} {2015})}\BibitemShut {NoStop}%
\bibitem [{\citenamefont {Feibelman}(2001)}]{feibelman_2001}%
  \BibitemOpen
  \bibfield  {author} {\bibinfo {author} {\bibfnamefont {P.~J.}\ \bibnamefont {Feibelman}},\ }\bibfield  {title} {\bibinfo {title} {Surface-diffusion mechanism versus electric field: Pt/pt(001)},\ }\href {https://doi.org/10.1103/PhysRevB.64.125403} {\bibfield  {journal} {\bibinfo  {journal} {Phys. Rev. B}\ }\textbf {\bibinfo {volume} {64}},\ \bibinfo {pages} {125403} (\bibinfo {year} {2001})}\BibitemShut {NoStop}%
\bibitem [{\citenamefont {Sanchez}\ \emph {et~al.}(2004)\citenamefont {Sanchez}, \citenamefont {Lozovoi},\ and\ \citenamefont {Alavi}}]{sanchez_2004}%
  \BibitemOpen
  \bibfield  {author} {\bibinfo {author} {\bibfnamefont {C.~G.}\ \bibnamefont {Sanchez}}, \bibinfo {author} {\bibfnamefont {A.~Y.}\ \bibnamefont {Lozovoi}},\ and\ \bibinfo {author} {\bibfnamefont {A.}~\bibnamefont {Alavi}},\ }\bibfield  {title} {\bibinfo {title} {Field-evaporation from first-principles},\ }\href {https://doi.org/10.1080/00268970410001727673} {\bibfield  {journal} {\bibinfo  {journal} {Molecular Physics}\ }\textbf {\bibinfo {volume} {102}},\ \bibinfo {pages} {1045} (\bibinfo {year} {2004})},\ \Eprint {https://arxiv.org/abs/https://doi.org/10.1080/00268970410001727673} {https://doi.org/10.1080/00268970410001727673} \BibitemShut {NoStop}%
\bibitem [{\citenamefont {Tamura}\ \emph {et~al.}(2012)\citenamefont {Tamura}, \citenamefont {Tsukada}, \citenamefont {McKenna}, \citenamefont {Shluger}, \citenamefont {Ohkubo},\ and\ \citenamefont {Hono}}]{tamura_2012}%
  \BibitemOpen
  \bibfield  {author} {\bibinfo {author} {\bibfnamefont {H.}~\bibnamefont {Tamura}}, \bibinfo {author} {\bibfnamefont {M.}~\bibnamefont {Tsukada}}, \bibinfo {author} {\bibfnamefont {K.~P.}\ \bibnamefont {McKenna}}, \bibinfo {author} {\bibfnamefont {A.~L.}\ \bibnamefont {Shluger}}, \bibinfo {author} {\bibfnamefont {T.}~\bibnamefont {Ohkubo}},\ and\ \bibinfo {author} {\bibfnamefont {K.}~\bibnamefont {Hono}},\ }\bibfield  {title} {\bibinfo {title} {Laser-assisted field evaporation from insulators triggered by photoinduced hole accumulation},\ }\href {https://doi.org/10.1103/PhysRevB.86.195430} {\bibfield  {journal} {\bibinfo  {journal} {Phys. Rev. B}\ }\textbf {\bibinfo {volume} {86}},\ \bibinfo {pages} {195430} (\bibinfo {year} {2012})}\BibitemShut {NoStop}%
\bibitem [{\citenamefont {Sundell}\ \emph {et~al.}(2019{\natexlab{a}})\citenamefont {Sundell}, \citenamefont {Hulander}, \citenamefont {Pihl},\ and\ \citenamefont {Andersson}}]{sundell_2019}%
  \BibitemOpen
  \bibfield  {author} {\bibinfo {author} {\bibfnamefont {G.}~\bibnamefont {Sundell}}, \bibinfo {author} {\bibfnamefont {M.}~\bibnamefont {Hulander}}, \bibinfo {author} {\bibfnamefont {A.}~\bibnamefont {Pihl}},\ and\ \bibinfo {author} {\bibfnamefont {M.}~\bibnamefont {Andersson}},\ }\bibfield  {title} {\bibinfo {title} {Atom probe tomography for 3d structural and chemical analysis of individual proteins},\ }\href@noop {} {\bibfield  {journal} {\bibinfo  {journal} {Small}\ }\textbf {\bibinfo {volume} {15}},\ \bibinfo {pages} {1900316} (\bibinfo {year} {2019}{\natexlab{a}})}\BibitemShut {NoStop}%
\bibitem [{\citenamefont {Hwang}\ \emph {et~al.}(2019)\citenamefont {Hwang}, \citenamefont {Park}, \citenamefont {Mun}, \citenamefont {Cho}, \citenamefont {Nam},\ and\ \citenamefont {Kim}}]{hwang2019generation}%
  \BibitemOpen
  \bibfield  {author} {\bibinfo {author} {\bibfnamefont {S.~I.}\ \bibnamefont {Hwang}}, \bibinfo {author} {\bibfnamefont {S.~B.}\ \bibnamefont {Park}}, \bibinfo {author} {\bibfnamefont {J.}~\bibnamefont {Mun}}, \bibinfo {author} {\bibfnamefont {W.}~\bibnamefont {Cho}}, \bibinfo {author} {\bibfnamefont {C.~H.}\ \bibnamefont {Nam}},\ and\ \bibinfo {author} {\bibfnamefont {K.~T.}\ \bibnamefont {Kim}},\ }\bibfield  {title} {\bibinfo {title} {Generation of a single-cycle pulse using a two-stage compressor and its temporal characterization using a tunnelling ionization method},\ }\href@noop {} {\bibfield  {journal} {\bibinfo  {journal} {Scientific reports}\ }\textbf {\bibinfo {volume} {9}},\ \bibinfo {pages} {1613} (\bibinfo {year} {2019})}\BibitemShut {NoStop}%
\bibitem [{\citenamefont {Blum}\ \emph {et~al.}(2016)\citenamefont {Blum}, \citenamefont {Cuvilly},\ and\ \citenamefont {Lefebvre-Ulrikson}}]{blum2016atom}%
  \BibitemOpen
  \bibfield  {author} {\bibinfo {author} {\bibfnamefont {I.}~\bibnamefont {Blum}}, \bibinfo {author} {\bibfnamefont {F.}~\bibnamefont {Cuvilly}},\ and\ \bibinfo {author} {\bibfnamefont {W.}~\bibnamefont {Lefebvre-Ulrikson}},\ }\bibfield  {title} {\bibinfo {title} {Atom probe sample preparation},\ }in\ \href@noop {} {\emph {\bibinfo {booktitle} {Atom Probe Tomography}}}\ (\bibinfo  {publisher} {Elsevier},\ \bibinfo {year} {2016})\ pp.\ \bibinfo {pages} {97--121}\BibitemShut {NoStop}%
\bibitem [{\citenamefont {Sundell}\ \emph {et~al.}(2019{\natexlab{b}})\citenamefont {Sundell}, \citenamefont {Hulander}, \citenamefont {Pihl},\ and\ \citenamefont {Andersson}}]{sundell2019atom}%
  \BibitemOpen
  \bibfield  {author} {\bibinfo {author} {\bibfnamefont {G.}~\bibnamefont {Sundell}}, \bibinfo {author} {\bibfnamefont {M.}~\bibnamefont {Hulander}}, \bibinfo {author} {\bibfnamefont {A.}~\bibnamefont {Pihl}},\ and\ \bibinfo {author} {\bibfnamefont {M.}~\bibnamefont {Andersson}},\ }\bibfield  {title} {\bibinfo {title} {Atom probe tomography for 3d structural and chemical analysis of individual proteins},\ }\href@noop {} {\bibfield  {journal} {\bibinfo  {journal} {Small}\ }\textbf {\bibinfo {volume} {15}},\ \bibinfo {pages} {1900316} (\bibinfo {year} {2019}{\natexlab{b}})}\BibitemShut {NoStop}%
\bibitem [{\citenamefont {Novi~Inverardi}\ \emph {et~al.}(2023)\citenamefont {Novi~Inverardi}, \citenamefont {Carnovale}, \citenamefont {Petrolli}, \citenamefont {Taioli},\ and\ \citenamefont {Lattanzi}}]{noviinverardi_2023}%
  \BibitemOpen
  \bibfield  {author} {\bibinfo {author} {\bibfnamefont {G.}~\bibnamefont {Novi~Inverardi}}, \bibinfo {author} {\bibfnamefont {F.}~\bibnamefont {Carnovale}}, \bibinfo {author} {\bibfnamefont {L.}~\bibnamefont {Petrolli}}, \bibinfo {author} {\bibfnamefont {S.}~\bibnamefont {Taioli}},\ and\ \bibinfo {author} {\bibfnamefont {G.}~\bibnamefont {Lattanzi}},\ }\bibfield  {title} {\bibinfo {title} {Silica in silico: A molecular dynamics characterization of the early stages of protein embedding for atom probe tomography},\ }\href {https://doi.org/10.3390/biophysica3020018} {\bibfield  {journal} {\bibinfo  {journal} {Biophysica}\ }\textbf {\bibinfo {volume} {3}},\ \bibinfo {pages} {276} (\bibinfo {year} {2023})}\BibitemShut {NoStop}%
\bibitem [{\citenamefont {Thompson}\ \emph {et~al.}(2007)\citenamefont {Thompson}, \citenamefont {Flaitz}, \citenamefont {Ronsheim}, \citenamefont {Larson},\ and\ \citenamefont {Kelly}}]{thompson2007imaging}%
  \BibitemOpen
  \bibfield  {author} {\bibinfo {author} {\bibfnamefont {K.}~\bibnamefont {Thompson}}, \bibinfo {author} {\bibfnamefont {P.~L.}\ \bibnamefont {Flaitz}}, \bibinfo {author} {\bibfnamefont {P.}~\bibnamefont {Ronsheim}}, \bibinfo {author} {\bibfnamefont {D.~J.}\ \bibnamefont {Larson}},\ and\ \bibinfo {author} {\bibfnamefont {T.~F.}\ \bibnamefont {Kelly}},\ }\bibfield  {title} {\bibinfo {title} {Imaging of arsenic cottrell atmospheres around silicon defects by three-dimensional atom probe tomography},\ }\href@noop {} {\bibfield  {journal} {\bibinfo  {journal} {Science}\ }\textbf {\bibinfo {volume} {317}},\ \bibinfo {pages} {1370} (\bibinfo {year} {2007})}\BibitemShut {NoStop}%
\bibitem [{\citenamefont {Blavette}\ \emph {et~al.}(1993)\citenamefont {Blavette}, \citenamefont {Bostel}, \citenamefont {Sarrau}, \citenamefont {Deconihout},\ and\ \citenamefont {Menand}}]{blavette1993atom}%
  \BibitemOpen
  \bibfield  {author} {\bibinfo {author} {\bibfnamefont {D.}~\bibnamefont {Blavette}}, \bibinfo {author} {\bibfnamefont {A.}~\bibnamefont {Bostel}}, \bibinfo {author} {\bibfnamefont {J.-M.}\ \bibnamefont {Sarrau}}, \bibinfo {author} {\bibfnamefont {B.}~\bibnamefont {Deconihout}},\ and\ \bibinfo {author} {\bibfnamefont {A.}~\bibnamefont {Menand}},\ }\bibfield  {title} {\bibinfo {title} {An atom probe for three-dimensional tomography},\ }\href@noop {} {\bibfield  {journal} {\bibinfo  {journal} {Nature}\ }\textbf {\bibinfo {volume} {363}},\ \bibinfo {pages} {432} (\bibinfo {year} {1993})}\BibitemShut {NoStop}%
\bibitem [{\citenamefont {Bartel}\ \emph {et~al.}(2005)\citenamefont {Bartel}, \citenamefont {Gaal}, \citenamefont {Reimann}, \citenamefont {Woerner},\ and\ \citenamefont {Elsaesser}}]{bartel2005generation}%
  \BibitemOpen
  \bibfield  {author} {\bibinfo {author} {\bibfnamefont {T.}~\bibnamefont {Bartel}}, \bibinfo {author} {\bibfnamefont {P.}~\bibnamefont {Gaal}}, \bibinfo {author} {\bibfnamefont {K.}~\bibnamefont {Reimann}}, \bibinfo {author} {\bibfnamefont {M.}~\bibnamefont {Woerner}},\ and\ \bibinfo {author} {\bibfnamefont {T.}~\bibnamefont {Elsaesser}},\ }\bibfield  {title} {\bibinfo {title} {Generation of single-cycle thz transients with high electric-field amplitudes},\ }\href@noop {} {\bibfield  {journal} {\bibinfo  {journal} {Optics Letters}\ }\textbf {\bibinfo {volume} {30}},\ \bibinfo {pages} {2805} (\bibinfo {year} {2005})}\BibitemShut {NoStop}%
\bibitem [{\citenamefont {Kim}\ \emph {et~al.}(2007)\citenamefont {Kim}, \citenamefont {Glownia}, \citenamefont {Taylor},\ and\ \citenamefont {Rodriguez}}]{kim2007terahertz}%
  \BibitemOpen
  \bibfield  {author} {\bibinfo {author} {\bibfnamefont {K.-Y.}\ \bibnamefont {Kim}}, \bibinfo {author} {\bibfnamefont {J.~H.}\ \bibnamefont {Glownia}}, \bibinfo {author} {\bibfnamefont {A.~J.}\ \bibnamefont {Taylor}},\ and\ \bibinfo {author} {\bibfnamefont {G.}~\bibnamefont {Rodriguez}},\ }\bibfield  {title} {\bibinfo {title} {Terahertz emission from ultrafast ionizing air in symmetry-broken laser fields},\ }\href@noop {} {\bibfield  {journal} {\bibinfo  {journal} {Optics express}\ }\textbf {\bibinfo {volume} {15}},\ \bibinfo {pages} {4577} (\bibinfo {year} {2007})}\BibitemShut {NoStop}%
\bibitem [{\citenamefont {Houard}\ \emph {et~al.}(2020)\citenamefont {Houard}, \citenamefont {Arnoldi}, \citenamefont {Ayoub}, \citenamefont {Hideur},\ and\ \citenamefont {Vella}}]{houard2020}%
  \BibitemOpen
  \bibfield  {author} {\bibinfo {author} {\bibfnamefont {J.}~\bibnamefont {Houard}}, \bibinfo {author} {\bibfnamefont {L.}~\bibnamefont {Arnoldi}}, \bibinfo {author} {\bibfnamefont {A.}~\bibnamefont {Ayoub}}, \bibinfo {author} {\bibfnamefont {A.}~\bibnamefont {Hideur}},\ and\ \bibinfo {author} {\bibfnamefont {A.}~\bibnamefont {Vella}},\ }\bibfield  {title} {\bibinfo {title} {Nanotip response to monocycle terahertz pulses},\ }\href@noop {} {\bibfield  {journal} {\bibinfo  {journal} {Applied Physics Letters}\ }\textbf {\bibinfo {volume} {117}} (\bibinfo {year} {2020})}\BibitemShut {NoStop}%
\bibitem [{\citenamefont {Costa}\ \emph {et~al.}(2012)\citenamefont {Costa}, \citenamefont {Wang}, \citenamefont {Duguay}, \citenamefont {Bostel}, \citenamefont {Blavette},\ and\ \citenamefont {Deconihout}}]{costa2012advance}%
  \BibitemOpen
  \bibfield  {author} {\bibinfo {author} {\bibfnamefont {G.~D.}\ \bibnamefont {Costa}}, \bibinfo {author} {\bibfnamefont {H.}~\bibnamefont {Wang}}, \bibinfo {author} {\bibfnamefont {S.}~\bibnamefont {Duguay}}, \bibinfo {author} {\bibfnamefont {A.}~\bibnamefont {Bostel}}, \bibinfo {author} {\bibfnamefont {D.}~\bibnamefont {Blavette}},\ and\ \bibinfo {author} {\bibfnamefont {B.}~\bibnamefont {Deconihout}},\ }\bibfield  {title} {\bibinfo {title} {Advance in multi-hit detection and quantization in atom probe tomography},\ }\href@noop {} {\bibfield  {journal} {\bibinfo  {journal} {Review of Scientific Instruments}\ }\textbf {\bibinfo {volume} {83}} (\bibinfo {year} {2012})}\BibitemShut {NoStop}%
\bibitem [{\citenamefont {Bitzek}\ \emph {et~al.}(2006)\citenamefont {Bitzek}, \citenamefont {Koskinen}, \citenamefont {G\"ahler}, \citenamefont {Moseler},\ and\ \citenamefont {Gumbsch}}]{Bitzek_2006}%
  \BibitemOpen
  \bibfield  {author} {\bibinfo {author} {\bibfnamefont {E.}~\bibnamefont {Bitzek}}, \bibinfo {author} {\bibfnamefont {P.}~\bibnamefont {Koskinen}}, \bibinfo {author} {\bibfnamefont {F.}~\bibnamefont {G\"ahler}}, \bibinfo {author} {\bibfnamefont {M.}~\bibnamefont {Moseler}},\ and\ \bibinfo {author} {\bibfnamefont {P.}~\bibnamefont {Gumbsch}},\ }\bibfield  {title} {\bibinfo {title} {Structural relaxation made simple},\ }\href {https://doi.org/10.1103/PhysRevLett.97.170201} {\bibfield  {journal} {\bibinfo  {journal} {Phys. Rev. Lett.}\ }\textbf {\bibinfo {volume} {97}},\ \bibinfo {pages} {170201} (\bibinfo {year} {2006})}\BibitemShut {NoStop}%
\bibitem [{\citenamefont {Umari}\ and\ \citenamefont {Pasquarello}(2002)}]{umari_2002}%
  \BibitemOpen
  \bibfield  {author} {\bibinfo {author} {\bibfnamefont {P.}~\bibnamefont {Umari}}\ and\ \bibinfo {author} {\bibfnamefont {A.}~\bibnamefont {Pasquarello}},\ }\bibfield  {title} {\bibinfo {title} {Ab initio molecular dynamics in a finite homogeneous electric field},\ }\href {https://doi.org/10.1103/PhysRevLett.89.157602} {\bibfield  {journal} {\bibinfo  {journal} {Phys. Rev. Lett.}\ }\textbf {\bibinfo {volume} {89}},\ \bibinfo {pages} {157602} (\bibinfo {year} {2002})}\BibitemShut {NoStop}%
\bibitem [{\citenamefont {Perdew}\ and\ \citenamefont {Wang}(1992)}]{perdew_1992}%
  \BibitemOpen
  \bibfield  {author} {\bibinfo {author} {\bibfnamefont {J.~P.}\ \bibnamefont {Perdew}}\ and\ \bibinfo {author} {\bibfnamefont {Y.}~\bibnamefont {Wang}},\ }\bibfield  {title} {\bibinfo {title} {Accurate and simple analytic representation of the electron-gas correlation energy},\ }\href {https://doi.org/10.1103/PhysRevB.45.13244} {\bibfield  {journal} {\bibinfo  {journal} {Phys. Rev. B}\ }\textbf {\bibinfo {volume} {45}},\ \bibinfo {pages} {13244} (\bibinfo {year} {1992})}\BibitemShut {NoStop}%
\bibitem [{\citenamefont {Marquis}\ \emph {et~al.}(2010)\citenamefont {Marquis}, \citenamefont {Yahya}, \citenamefont {Larson}, \citenamefont {Miller},\ and\ \citenamefont {Todd}}]{marquis2010probing}%
  \BibitemOpen
  \bibfield  {author} {\bibinfo {author} {\bibfnamefont {E.~A.}\ \bibnamefont {Marquis}}, \bibinfo {author} {\bibfnamefont {N.~A.}\ \bibnamefont {Yahya}}, \bibinfo {author} {\bibfnamefont {D.~J.}\ \bibnamefont {Larson}}, \bibinfo {author} {\bibfnamefont {M.~K.}\ \bibnamefont {Miller}},\ and\ \bibinfo {author} {\bibfnamefont {R.~I.}\ \bibnamefont {Todd}},\ }\bibfield  {title} {\bibinfo {title} {Probing the improbable: Imaging c atoms in alumina},\ }\href@noop {} {\bibfield  {journal} {\bibinfo  {journal} {Materials Today}\ }\textbf {\bibinfo {volume} {13}},\ \bibinfo {pages} {34} (\bibinfo {year} {2010})}\BibitemShut {NoStop}%
\bibitem [{\citenamefont {Vella}\ \emph {et~al.}(2011)\citenamefont {Vella}, \citenamefont {Mazumder}, \citenamefont {Da~Costa},\ and\ \citenamefont {Deconihout}}]{vella2011field}%
  \BibitemOpen
  \bibfield  {author} {\bibinfo {author} {\bibfnamefont {A.}~\bibnamefont {Vella}}, \bibinfo {author} {\bibfnamefont {B.}~\bibnamefont {Mazumder}}, \bibinfo {author} {\bibfnamefont {G.}~\bibnamefont {Da~Costa}},\ and\ \bibinfo {author} {\bibfnamefont {B.}~\bibnamefont {Deconihout}},\ }\bibfield  {title} {\bibinfo {title} {Field evaporation mechanism of bulk oxides under ultra fast laser illumination},\ }\href@noop {} {\bibfield  {journal} {\bibinfo  {journal} {Journal of Applied Physics}\ }\textbf {\bibinfo {volume} {110}} (\bibinfo {year} {2011})}\BibitemShut {NoStop}%
\bibitem [{\citenamefont {Constanda}\ \emph {et~al.}(2016)\citenamefont {Constanda}, \citenamefont {Doty},\ and\ \citenamefont {Hamill}}]{BEM}%
  \BibitemOpen
  \bibfield  {author} {\bibinfo {author} {\bibfnamefont {C.}~\bibnamefont {Constanda}}, \bibinfo {author} {\bibfnamefont {D.}~\bibnamefont {Doty}},\ and\ \bibinfo {author} {\bibfnamefont {W.}~\bibnamefont {Hamill}},\ }\bibfield  {title} {\bibinfo {title} {Boundary integral equation methods and numerical solutions},\ }\href@noop {} {\bibfield  {journal} {\bibinfo  {journal} {Dev. Math}\ }\textbf {\bibinfo {volume} {35}} (\bibinfo {year} {2016})}\BibitemShut {NoStop}%
\bibitem [{\citenamefont {Andrade}\ \emph {et~al.}(2015)\citenamefont {Andrade}, \citenamefont {Strubbe}, \citenamefont {De~Giovannini}, \citenamefont {Larsen}, \citenamefont {Oliveira}, \citenamefont {Alberdi-Rodriguez}, \citenamefont {Varas}, \citenamefont {Theophilou}, \citenamefont {Helbig}, \citenamefont {Verstraete}, \citenamefont {Stella}, \citenamefont {Nogueira}, \citenamefont {Aspuru-Guzik}, \citenamefont {Castro}, \citenamefont {Marques},\ and\ \citenamefont {Rubio}}]{octopus_2015}%
  \BibitemOpen
  \bibfield  {author} {\bibinfo {author} {\bibfnamefont {X.}~\bibnamefont {Andrade}}, \bibinfo {author} {\bibfnamefont {D.}~\bibnamefont {Strubbe}}, \bibinfo {author} {\bibfnamefont {U.}~\bibnamefont {De~Giovannini}}, \bibinfo {author} {\bibfnamefont {A.~H.}\ \bibnamefont {Larsen}}, \bibinfo {author} {\bibfnamefont {M.~J.~T.}\ \bibnamefont {Oliveira}}, \bibinfo {author} {\bibfnamefont {J.}~\bibnamefont {Alberdi-Rodriguez}}, \bibinfo {author} {\bibfnamefont {A.}~\bibnamefont {Varas}}, \bibinfo {author} {\bibfnamefont {I.}~\bibnamefont {Theophilou}}, \bibinfo {author} {\bibfnamefont {N.}~\bibnamefont {Helbig}}, \bibinfo {author} {\bibfnamefont {M.~J.}\ \bibnamefont {Verstraete}}, \bibinfo {author} {\bibfnamefont {L.}~\bibnamefont {Stella}}, \bibinfo {author} {\bibfnamefont {F.}~\bibnamefont {Nogueira}}, \bibinfo {author} {\bibfnamefont {A.}~\bibnamefont {Aspuru-Guzik}}, \bibinfo {author} {\bibfnamefont {A.}~\bibnamefont {Castro}}, \bibinfo {author} {\bibfnamefont {M.~A.~L.}\ \bibnamefont {Marques}},\ and\
  \bibinfo {author} {\bibfnamefont {A.}~\bibnamefont {Rubio}},\ }\bibfield  {title} {\bibinfo {title} {Real-space grids and the octopus code as tools for the development of new simulation approaches for electronic systems},\ }\href {https://doi.org/10.1039/C5CP00351B} {\bibfield  {journal} {\bibinfo  {journal} {Phys. Chem. Chem. Phys.}\ }\textbf {\bibinfo {volume} {17}},\ \bibinfo {pages} {31371} (\bibinfo {year} {2015})}\BibitemShut {NoStop}%
\bibitem [{\citenamefont {Tancogne-Dejean}\ \emph {et~al.}(2020)\citenamefont {Tancogne-Dejean}, \citenamefont {Oliveira}, \citenamefont {Andrade}, \citenamefont {Appel}, \citenamefont {Borca}, \citenamefont {Le~Breton}, \citenamefont {Buchholz}, \citenamefont {Castro}, \citenamefont {Corni}, \citenamefont {Correa}, \citenamefont {De~Giovannini}, \citenamefont {Delgado}, \citenamefont {Eich}, \citenamefont {Flick}, \citenamefont {Gil}, \citenamefont {Gomez}, \citenamefont {Helbig}, \citenamefont {H{\"u}bener}, \citenamefont {Jest{\"a}dt}, \citenamefont {Jornet-Somoza}, \citenamefont {Larsen}, \citenamefont {Lebedeva}, \citenamefont {L{\"u}ders}, \citenamefont {Marques}, \citenamefont {Ohlmann}, \citenamefont {Pipolo}, \citenamefont {Rampp}, \citenamefont {Rozzi}, \citenamefont {Strubbe}, \citenamefont {Sato}, \citenamefont {Sch{\"a}fer}, \citenamefont {Theophilou}, \citenamefont {Welden},\ and\ \citenamefont {Rubio}}]{octopus_2020}%
  \BibitemOpen
  \bibfield  {author} {\bibinfo {author} {\bibfnamefont {N.}~\bibnamefont {Tancogne-Dejean}}, \bibinfo {author} {\bibfnamefont {M.~J.~T.}\ \bibnamefont {Oliveira}}, \bibinfo {author} {\bibfnamefont {X.}~\bibnamefont {Andrade}}, \bibinfo {author} {\bibfnamefont {H.}~\bibnamefont {Appel}}, \bibinfo {author} {\bibfnamefont {C.~H.}\ \bibnamefont {Borca}}, \bibinfo {author} {\bibfnamefont {G.}~\bibnamefont {Le~Breton}}, \bibinfo {author} {\bibfnamefont {F.}~\bibnamefont {Buchholz}}, \bibinfo {author} {\bibfnamefont {A.}~\bibnamefont {Castro}}, \bibinfo {author} {\bibfnamefont {S.}~\bibnamefont {Corni}}, \bibinfo {author} {\bibfnamefont {A.~A.}\ \bibnamefont {Correa}}, \bibinfo {author} {\bibfnamefont {U.}~\bibnamefont {De~Giovannini}}, \bibinfo {author} {\bibfnamefont {A.}~\bibnamefont {Delgado}}, \bibinfo {author} {\bibfnamefont {F.~G.}\ \bibnamefont {Eich}}, \bibinfo {author} {\bibfnamefont {J.}~\bibnamefont {Flick}}, \bibinfo {author} {\bibfnamefont {G.}~\bibnamefont {Gil}}, \bibinfo {author} {\bibfnamefont
  {A.}~\bibnamefont {Gomez}}, \bibinfo {author} {\bibfnamefont {N.}~\bibnamefont {Helbig}}, \bibinfo {author} {\bibfnamefont {H.}~\bibnamefont {H{\"u}bener}}, \bibinfo {author} {\bibfnamefont {R.}~\bibnamefont {Jest{\"a}dt}}, \bibinfo {author} {\bibfnamefont {J.}~\bibnamefont {Jornet-Somoza}}, \bibinfo {author} {\bibfnamefont {A.~H.}\ \bibnamefont {Larsen}}, \bibinfo {author} {\bibfnamefont {I.~V.}\ \bibnamefont {Lebedeva}}, \bibinfo {author} {\bibfnamefont {M.}~\bibnamefont {L{\"u}ders}}, \bibinfo {author} {\bibfnamefont {M.~A.~L.}\ \bibnamefont {Marques}}, \bibinfo {author} {\bibfnamefont {S.~T.}\ \bibnamefont {Ohlmann}}, \bibinfo {author} {\bibfnamefont {S.}~\bibnamefont {Pipolo}}, \bibinfo {author} {\bibfnamefont {M.}~\bibnamefont {Rampp}}, \bibinfo {author} {\bibfnamefont {C.~A.}\ \bibnamefont {Rozzi}}, \bibinfo {author} {\bibfnamefont {D.~A.}\ \bibnamefont {Strubbe}}, \bibinfo {author} {\bibfnamefont {S.~A.}\ \bibnamefont {Sato}}, \bibinfo {author} {\bibfnamefont {C.}~\bibnamefont {Sch{\"a}fer}},
  \bibinfo {author} {\bibfnamefont {I.}~\bibnamefont {Theophilou}}, \bibinfo {author} {\bibfnamefont {A.}~\bibnamefont {Welden}},\ and\ \bibinfo {author} {\bibfnamefont {A.}~\bibnamefont {Rubio}},\ }\bibfield  {title} {\bibinfo {title} {{Octopus, a computational framework for exploring light-driven phenomena and quantum dynamics in extended and finite systems}},\ }\href {https://doi.org/10.1063/1.5142502} {\bibfield  {journal} {\bibinfo  {journal} {The Journal of Chemical Physics}\ }\textbf {\bibinfo {volume} {152}},\ \bibinfo {pages} {124119} (\bibinfo {year} {2020})},\ \Eprint {https://arxiv.org/abs/https://pubs.aip.org/aip/jcp/article-pdf/doi/10.1063/1.5142502/16712083/124119\_1\_online.pdf} {https://pubs.aip.org/aip/jcp/article-pdf/doi/10.1063/1.5142502/16712083/124119\_1\_online.pdf} \BibitemShut {NoStop}%
\bibitem [{\citenamefont {Belton}\ \emph {et~al.}(2012)\citenamefont {Belton}, \citenamefont {Deschaume},\ and\ \citenamefont {Perry}}]{belton_2012}%
  \BibitemOpen
  \bibfield  {author} {\bibinfo {author} {\bibfnamefont {D.}~\bibnamefont {Belton}}, \bibinfo {author} {\bibfnamefont {O.}~\bibnamefont {Deschaume}},\ and\ \bibinfo {author} {\bibfnamefont {C.}~\bibnamefont {Perry}},\ }\bibfield  {title} {\bibinfo {title} {An overview of the fundamentals of the chemistry of silica with relevance to biosilicification and technological advances},\ }\href {https://doi.org/doi:10.1111/j.1742-4658.2012.08531.x} {\bibfield  {journal} {\bibinfo  {journal} {FEBS J}\ }\textbf {\bibinfo {volume} {279}},\ \bibinfo {pages} {1710} (\bibinfo {year} {2012})}\BibitemShut {NoStop}%
\bibitem [{\citenamefont {Vella}\ \emph {et~al.}(2021{\natexlab{b}})\citenamefont {Vella}, \citenamefont {Houard}, \citenamefont {Arnoldi}, \citenamefont {Tang}, \citenamefont {Boudant}, \citenamefont {Ayoub}, \citenamefont {Normand}, \citenamefont {Da~Costa},\ and\ \citenamefont {Hideur}}]{vella_2021}%
  \BibitemOpen
  \bibfield  {author} {\bibinfo {author} {\bibfnamefont {A.}~\bibnamefont {Vella}}, \bibinfo {author} {\bibfnamefont {J.}~\bibnamefont {Houard}}, \bibinfo {author} {\bibfnamefont {L.}~\bibnamefont {Arnoldi}}, \bibinfo {author} {\bibfnamefont {M.}~\bibnamefont {Tang}}, \bibinfo {author} {\bibfnamefont {M.}~\bibnamefont {Boudant}}, \bibinfo {author} {\bibfnamefont {A.}~\bibnamefont {Ayoub}}, \bibinfo {author} {\bibfnamefont {A.}~\bibnamefont {Normand}}, \bibinfo {author} {\bibfnamefont {G.}~\bibnamefont {Da~Costa}},\ and\ \bibinfo {author} {\bibfnamefont {A.}~\bibnamefont {Hideur}},\ }\bibfield  {title} {\bibinfo {title} {High-resolution terahertz-driven atom probe tomography},\ }\href@noop {} {\bibfield  {journal} {\bibinfo  {journal} {Science Advances}\ }\textbf {\bibinfo {volume} {7}},\ \bibinfo {pages} {eabd7259} (\bibinfo {year} {2021}{\natexlab{b}})}\BibitemShut {NoStop}%
\end{thebibliography}%



\newpage

%
%
%




\begin{CJK*}{}{} 
\author{Matteo De Tullio,$^{1\ast}$ 
Giovanni Novi Inverardi,$^{2}$ 
Michella Karam,$^{1}$ 
Jonathan Houard,$^{1}$ 
Marc Ropitaux,$^{3}$ 
Ivan Blum,$^{1}$ 
Francesco Carnovale,$^{2}$ 
Gianluca Lattanzi,$^{4}$ 
Simone Taioli,$^{5}$ 
Gustav Eriksson,$^{6}$ 
Mats Hulander,$^{6}$ 
Martin Andersson,$^{6}$ 
Angela Vella$^{1\ast}$
Tommaso Morresi$^{5\ast}$\\~\\}
\affiliation{$^{1}$Universit{\'e} Rouen Normandie, INSA Rouen Normandie, CNRS, GPM UMR 6634, F-76000 Rouen, France\\
$^{2}$Department of Physics, University of Trento and European Centre for Theoretical Studies in Nuclear Physics and Related Areas (ECT*-FBK) and Trento Institute for Fundamental Physics and Applications (TIFPA-INFN), Trento, Italy\\
$^{3}$Universit{\'e} Rouen Normandie, GLYCOMEV UR4358, SFR Normandie V{\'e}g{\'e}tal FED 4277, Innovation Chimie Carnot, IRIB, F-76000 Rouen, France\\
$^{4}$Department of Physics, University of Trento and Trento Institute for Fundamental Physics and Applications (TIFPA-INFN), Trento, Italy\\
$^{5}$European Centre for Theoretical Studies in Nuclear Physics and Related Areas (ECT*-FBK) and Trento Institute for Fundamental Physics and Applications (TIFPA-INFN), Trento, Italy\\
$^{6}$Department of Chemistry and Chemical Engineering, Chalmers University of Technology Gothenburg 41296, Sweden
}%
\title{\LARGE{SUPPLEMENTAL INFORMATION}\\Evaporation of cations from non-conductive nano-samples using single-cycle THz pulses: an experimental and theoretical study}

\maketitle
\end{CJK*}
\setcounter{section}{0}
\setcounter{figure}{0}
\renewcommand{\theequation}{S.\arabic{equation}}
\renewcommand{\thefigure}{S.\arabic{figure}}
\renewcommand{\thetable}{S.\arabic{table}}
\renewcommand{\thesection}{S.\arabic{section}}
\renewcommand*{\citenumfont}[1]{S#1}
\renewcommand*{\bibnumfmt}[1]{[S#1]}

\section{1D MODEL USING UV FREQUENCIES}
\begin{figure}[hbt!]
\centering
\includegraphics[width=0.49\textwidth]{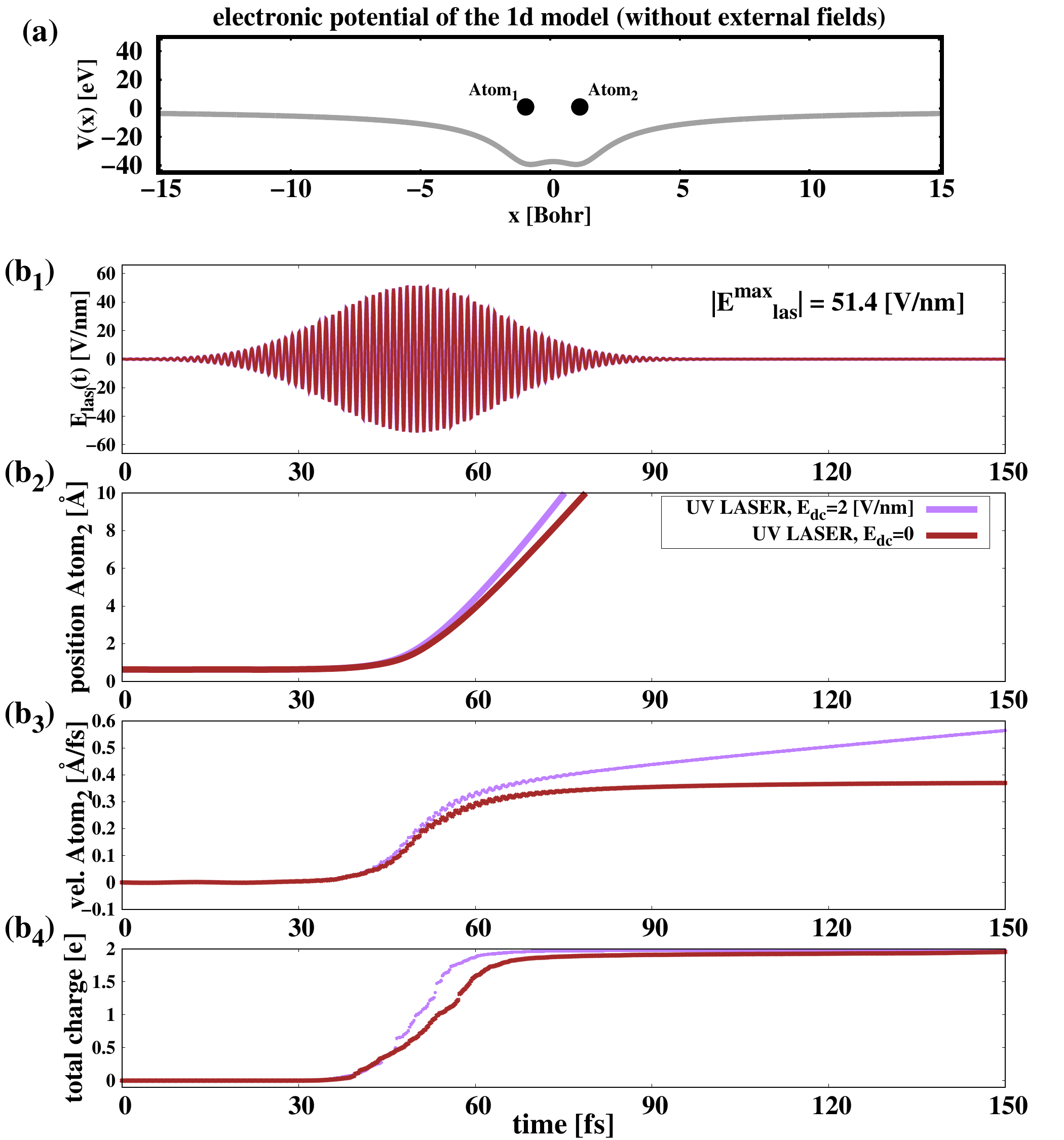}
\includegraphics[width=0.49\textwidth]{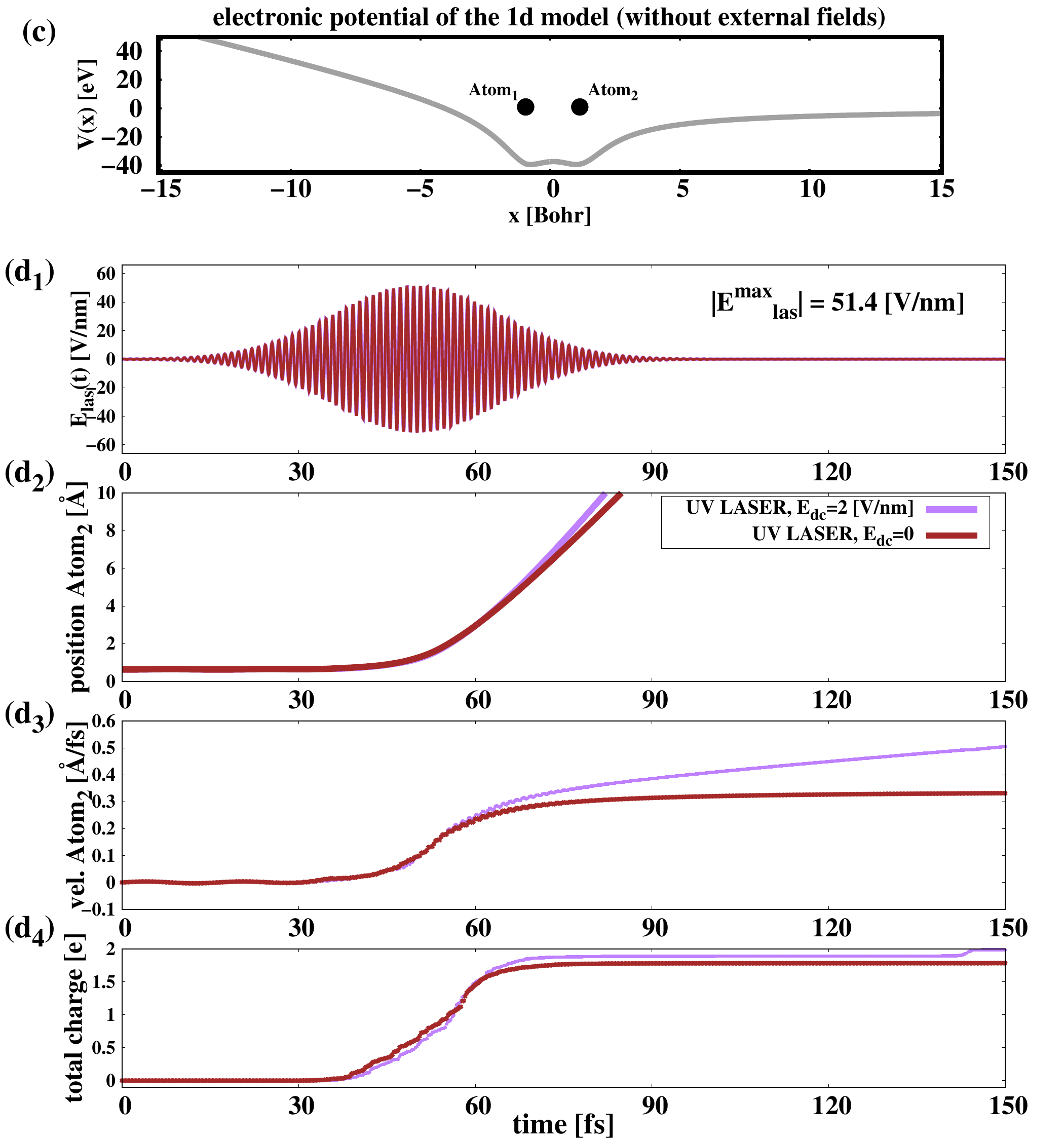}
\caption{Comparison of the results obtained with the two potentials in (a) and (c) for a fixed amplitude of a UV laser of frequency $\omega$=3.61 eV in (b$_1$) and (d$_1$). The maximum amplitude is 0.1 atomic units, as in Fig. 8 of the main text. In (b$_2$) and (d$_2$) the position of Atom$_2$ is shown as a function of time and in (b$_3$) and (d$_3$) its velocity in the case of laser plus static potential (purple line) and without external potential (brown line). Finally, the total charge inside the box is given in (b$_4$) and (d$_4$).
On the left side (for metal samples, (a),(b$_{1-4}$)) Atom$_2$ evaporates when a positive THz pulse is used (light green and dark green lines) due to the strong ionisation (b$_4$). On the right-hand side (for insulating materials, (c),(d$_{1-4}$)), exactly the same evaporation is observed when a linear potential barrier is added. In contrast to the THz pulses, the potential is no longer effective when using UV lasers, as the electrons can also be ejected from the surface (right-hand sides in (a) and (c)). \label{fig:SI}}
\end{figure}

\newpage
\section{1D MODEL USING A SINGLE-CYCLE INFRARED PULSE}
\begin{figure}[hbt!]
\centering
\includegraphics[width=0.49\textwidth]{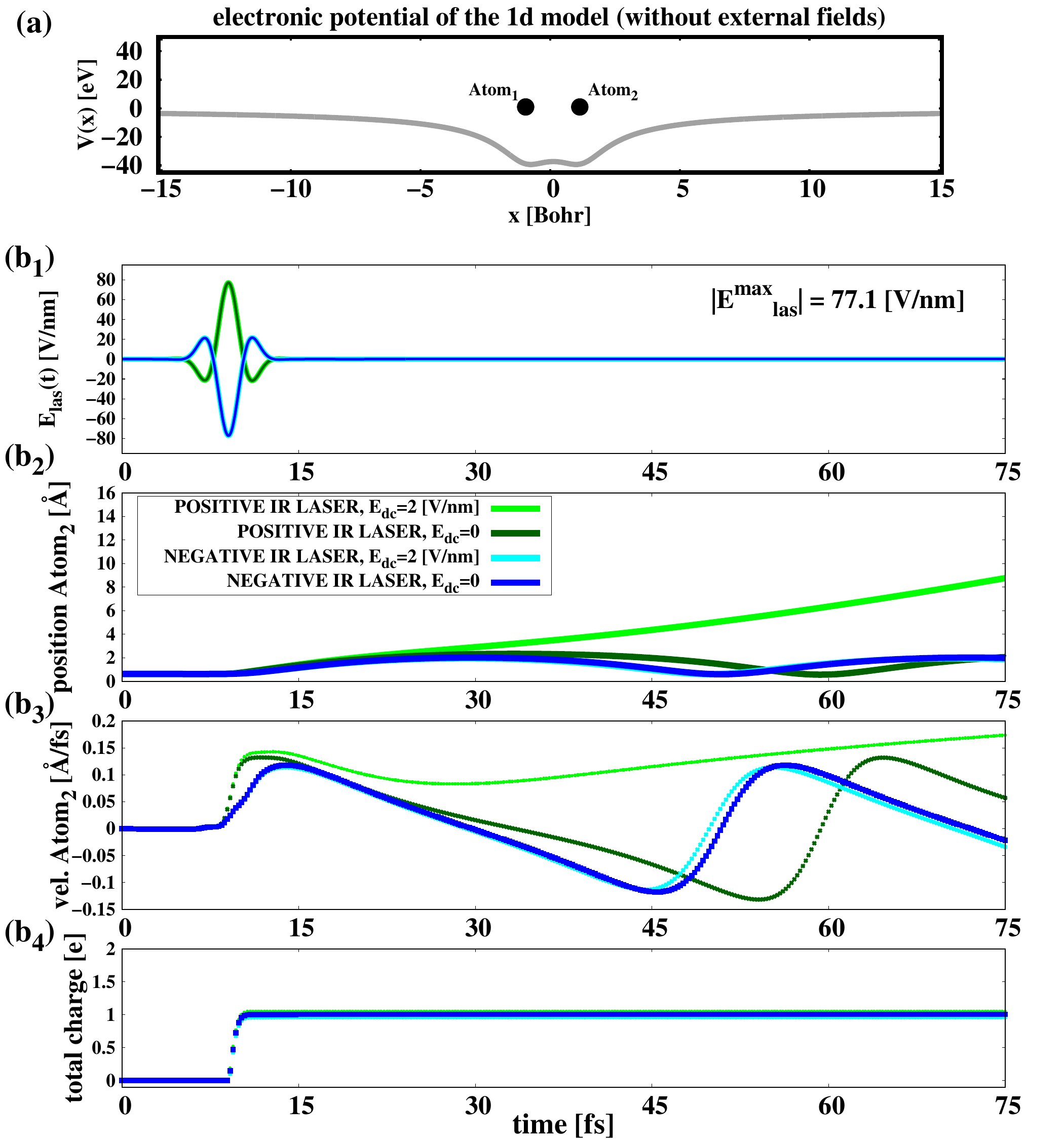}
\includegraphics[width=0.49\textwidth]{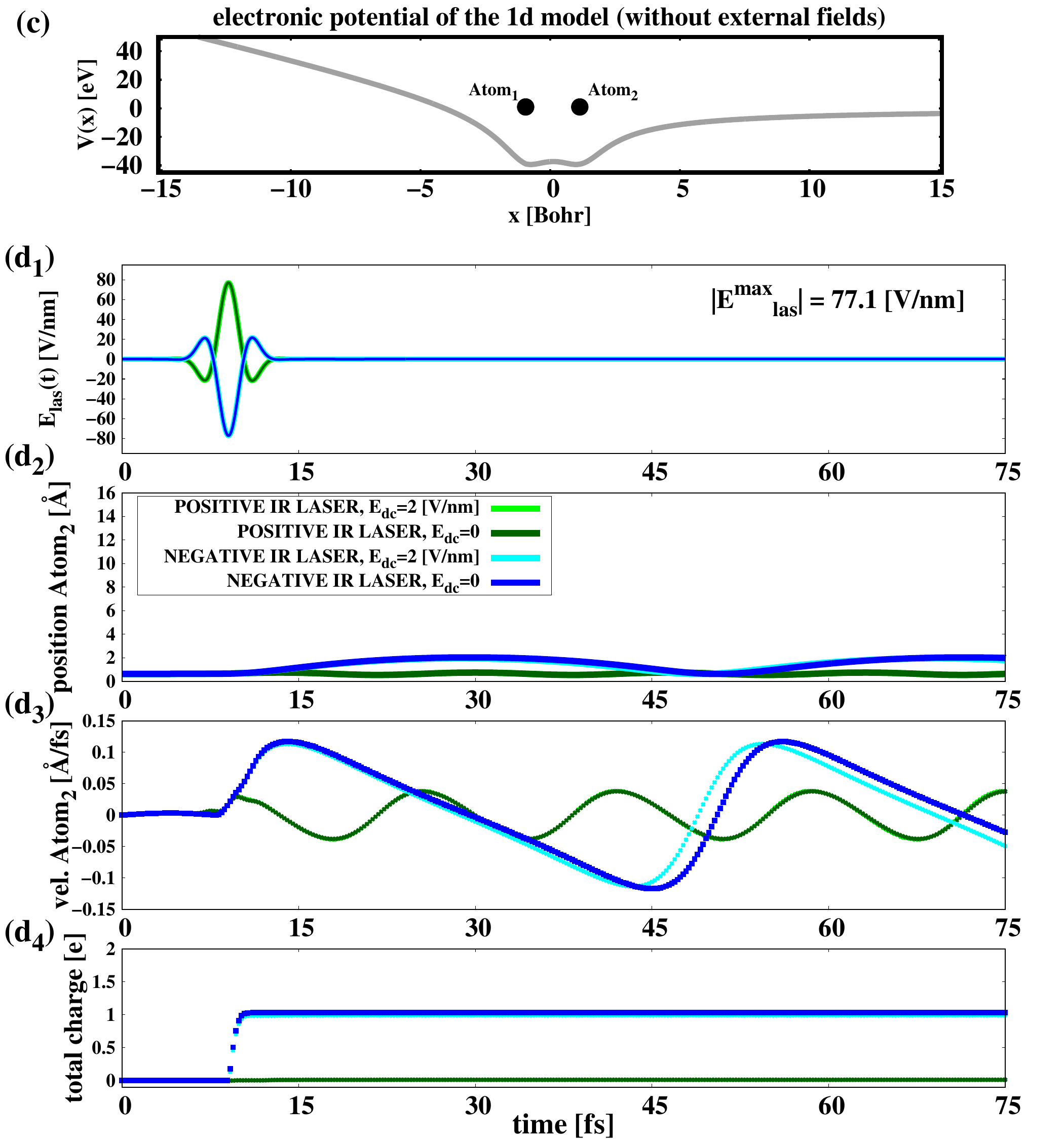}
\caption{Comparison of the results obtained with the two potentials in (a) and (c) for a fixed amplitude of a IR single-pulse of frequency $\omega$=0.799898 eV, corresponding to a wavelength of 1.55 $\mu m$, in (b$_1$) and (d$_1$). The maximum amplitude is 0.15 atomic units. The light and dark green curves correspond to the positive pulse, while the cyan and blue lines represent the negative pulse. In (b$_2$) and (d$_2$) the position of Atom$_2$ is shown as a function of time and in (b$_3$) and (d$_3$) its velocity in the different cases. Finally, (b$_4$) and (d$_4$) show the total charge inside the box.
On the left side (for metal-like samples, (a),(b$_{1-4}$)) Atom$_2$ evaporates when a positive pulse is used in conjunction with the static external field (light green line) due to the strong ionisation (b$_4$). On the right-hand side (for insulating-like materials, (c),(d$_{1-4}$)), evaporation does not take place. This is similar to the findings for THz pulses in Fig. 8 of the main text, although a larger pulse-amplitude is used in this case. \label{fig:SI_IR1}}
\end{figure}

\begin{figure}[hbt!]
\centering
\includegraphics[width=0.49\textwidth]{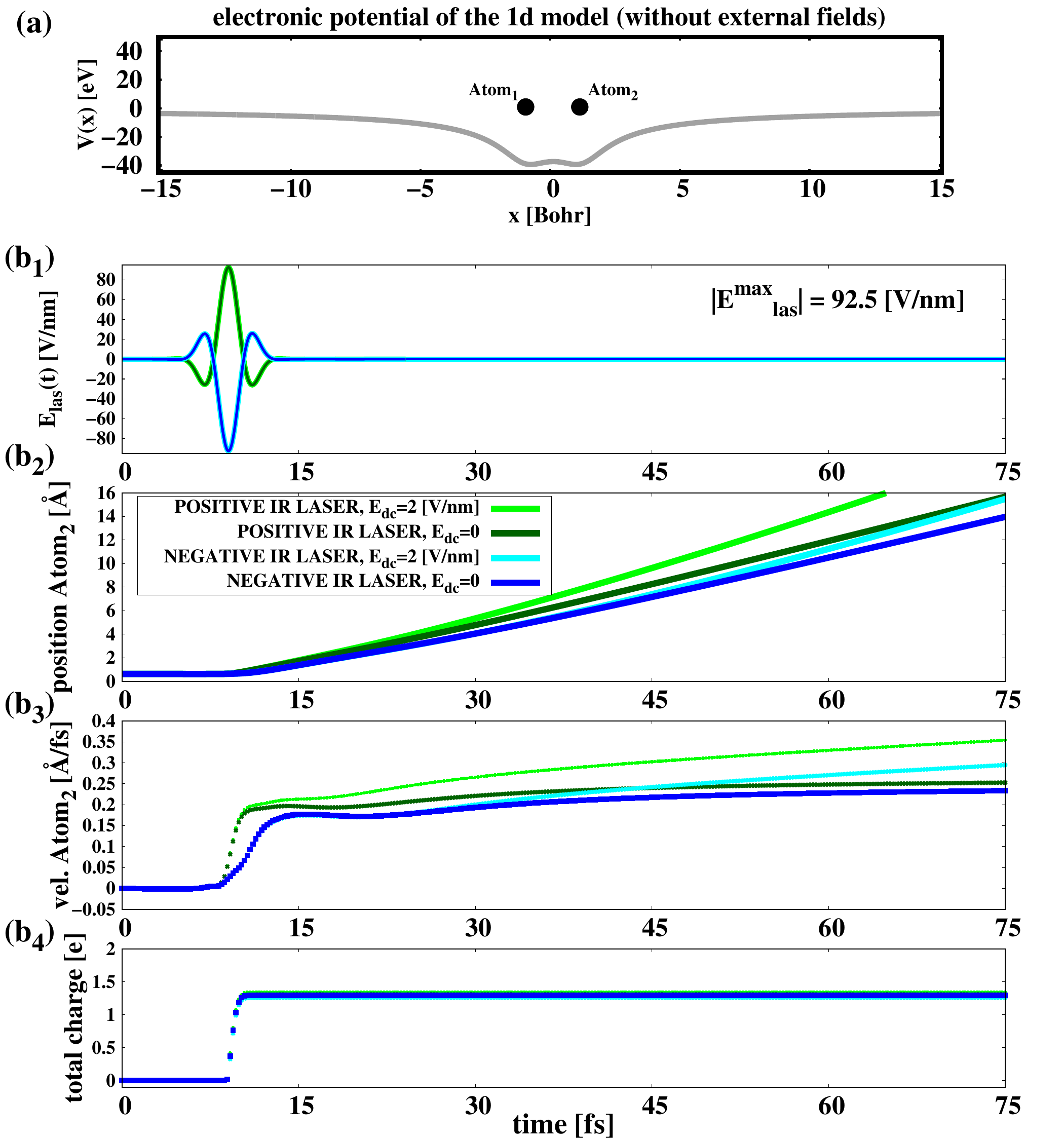}
\includegraphics[width=0.49\textwidth]{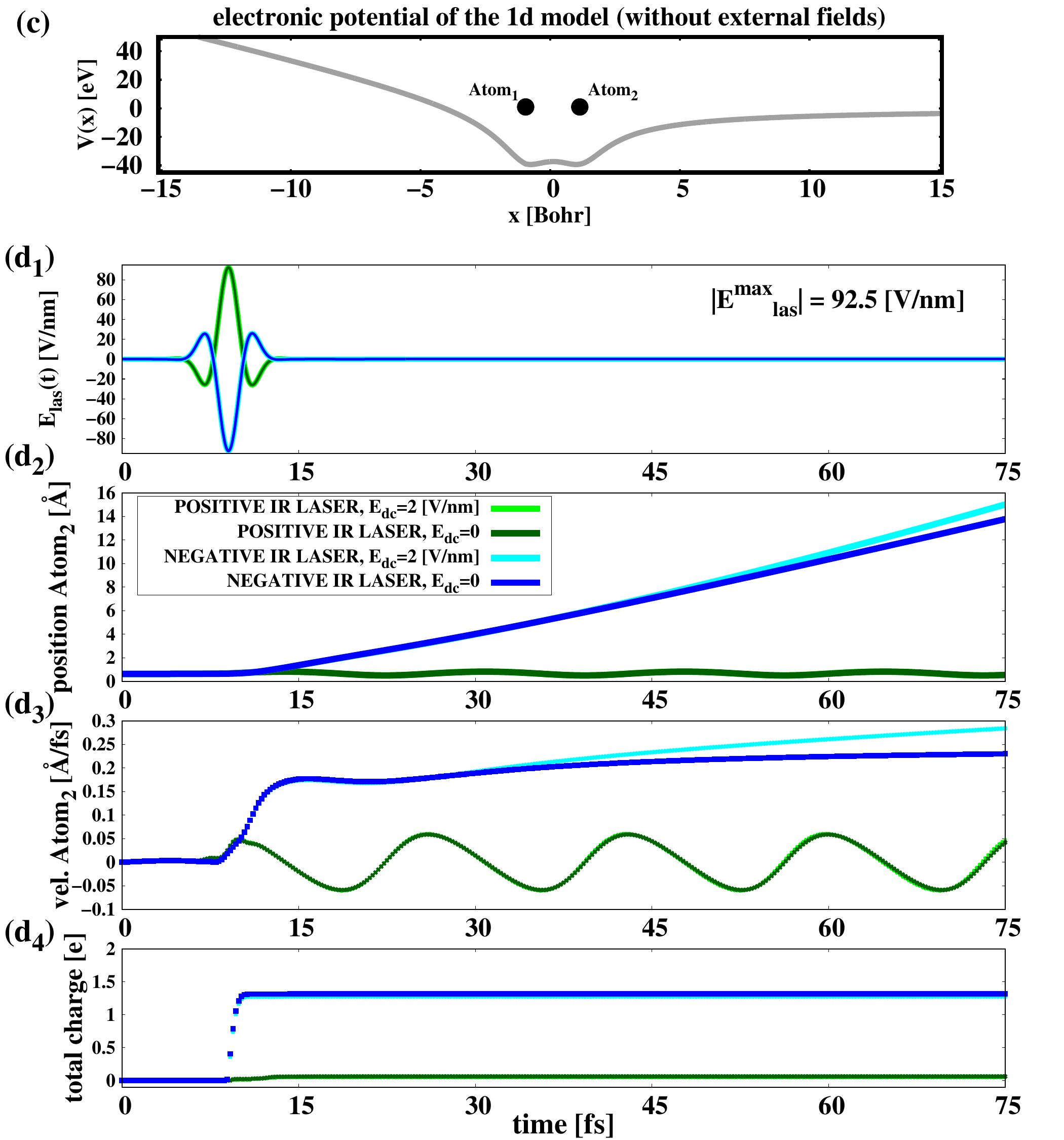}
\caption{Comparison of the results obtained with the two potentials in (a) and (c) for a fixed amplitude of a IR single-pulse of frequency $\omega$=0.799898 eV, corresponding to a wavelength of 1.55 $\mu m$, in (b$_1$) and (d$_1$). The maximum amplitude is 0.18 atomic units, larger than in Fig. \ref{fig:SI_IR1}. The light and dark green curves correspond to the positive pulse, while the cyan and blue lines represent the negative pulse. In (b$_2$) and (d$_2$) the position of Atom$_2$ is shown as a function of time and in (b$_3$) and (d$_3$) its velocity in the different cases. Finally, (b$_4$) and (d$_4$) show the total charge inside the box.
On the left side (for metal-like samples, (a),(b$_{1-4}$)) Atom$_2$ always evaporates due to the strong ionisation (b$_4$), which results to be the same for the two laser pulses. On the right-hand side (for insulating-like materials, (c),(d$_{1-4}$)), evaporation does not take place in the case of positive pulses. This resembles the behaviour using THz pulses in Fig. 9 of the main text, although a larger pulse-amplitude is used in this case. \label{fig:SI_IR2}}
\end{figure}

\end{document}